\documentclass{aa}

\usepackage{graphicx}
\usepackage{mwe}
\usepackage{caption}
\usepackage{subcaption}
\usepackage{float}
\usepackage{enumitem}
\usepackage{xcolor}
\usepackage[none]{hyphenat}
\usepackage{adjustbox}
\usepackage{xcolor, colortbl}
\usepackage{array, multirow, multicol}
\usepackage{amsmath}   
\usepackage{txfonts}
\usepackage{hyperref}
\usepackage{soul}

\usepackage{ulem}

\begin{document}

   \title{Multipolar magnetic configuration: a trace of post-mergers events in circumstellar disks around FS CMa Stars}
  \titlerunning{A trace of post-mergers events in circumstellar disks}

   \author{A. Moranchel-Basurto
          \inst{1},
          D. Kor\v{c}\'{a}kov\'{a}\inst{1},
          M. \v{C}emelji\'{c}\inst{2,3,4},
          R. O. Chametla\inst{1},
          F. J. S\'{a}nchez-Salcedo\inst{5} 
          \and
          I.~Bermejo-Lozano\inst{1}
          }
    \authorrunning{A. Moranchel-Basurto, et al.}
   
   \institute{Charles University, Faculty of Mathematics and Physics, Astronomical Institute. V Hole\v{s}ovi\v{c}k\'{a}ch 747/2, 180 00, Prague 8, CZ-746\,01 Opava, Czech Republic\\
    \email{a.moranchel@mff.cuni.cz}
    \and
    Nicolaus Copernicus Astronomical Center of the Polish Academy of Sciences, Bartycka 18, 00-716, Warsaw, Poland 
\and
    Nicolaus Copernicus Superior School, College of Astronomy and Natural Sciences, Gregorkiewicza 3, 87-100, Toru\'{n}, Poland
    \and
Research Centre for Computational Physics and Data Processing, Institute of Physics, Silesian University in Opava, Bezru\v{c}ovo n\'am.~13, CZ-746\,01 Opava, Czech Republic
    \and
    Instituto de Astronomía, Universidad Nacional Autónoma de Mexico,  A. P. 70-264, 04510, CDMX, México
}

   \date{Received XXX; accepted YYY}

 
  \abstract
   {
   Observations suggest that magnetic fields of disk-bearing stars may have non-dipolar configurations. However, the influence of these configurations on magnetospheric accretion remains poorly understood.
   }
    {We aim to simulate magnetospheric accretion incorporating non-dipolar and strong magnetic field. Our model is informed by observations of IRAS 17449+2320, a post-merger belonging to the group of FS CMa stars, which indicate a dominant dipolar magnetic field with an additional quadrupole component.
    }
   {Using the PLUTO code, we conduct 2.5-D non-ideal viscous-resistive ($\alpha_\nu=1$ and $\alpha_\mathrm{m}=1$) magnetohydrodynamical (MHD) simulations of star-disk magnetospheric interactions. We consider a thin accretion disk and strong stellar magnetic field ($B_\star= 6.2\mathrm{kG}$) under four configurations: pure dipole, pure quadrupole, dipole plus quadrupole, and dipole plus octupole. In the latter two cases, different magnetic polar strength ratios are explored.}
   {For asymmetric magnetic field configurations, we find that accretion exhibits funnel streams below the midplane, indicating the dominance of the quadrupolar and octupolar components. In contrast, in dipolar configurations, we observe the formation of two symmetrical funnels  with respect to the midplane. However, in the quadrupolar configuration, accretion is entirely confined to the disk midplane forming a cone-like pattern that leads to disk widening. Remarkably, the presence of a quadrupolar component gives rise to highly asymmetric substructures in the corona region.
   }
   {Multipolar stellar magnetic fields drive non-uniform accretion and lead to asymmetric density distributions in both the disk and corona. These results resemble observed features of some FS CMa post-mergers and Herbig Ae/Be stars, highlighting the critical role of magnetic field complexity in shaping circumstellar environments.}

   \keywords{star-disk interaction --
                accretion disks --
                magnetohydrodynamics--star formation -- Herbig Ae/Be stars
               }

   \maketitle
%

\section{Introduction}

Many astrophysical systems consist of a central magnetized star surrounded by an accretion disk. One of the most widely accepted magnetospheric accretion models to date is that of \cite{Ghosh78}, which assumes that near the star, magnetic fields are so strong that the plasma is forced to corotate with the
star. It relies on the balance between the spin-up torque on the star, arisen from the interaction with the region inside the corotation radius, and the spin-down torque from the region outside it. 
It includes field shearing due to the disk flow. In this model, the growth of the toroidal field component ($B_\phi$) is limited by reconnection, and it is assumed that a steady state is reached. 
Many of the predicted features were confirmed through numerical simulations, and new insights into the accretion process onto a star with a dipole field were revealed. For example, it was found that matter can fall directly onto the star's equator \citep{Mishra2023} or form an accretion column at a higher latitude, so-called funnel effect \citep[][]{Romanova_etal2002,Zanni2009,Cemeljic2019}, but it can also escape from the vicinity of the star in the form of a stellar wind or conical and axial outflow \citep{Zanni2013,Kotek2020,Cemeljic2023}. 

However, there are several observational results indicating that the configuration of the magnetic field of strongly magnetized stars may depart from the dipole one. For instance, \cite{bouvier06} argued that there are many signs of magnetospheric accretion in Classical T~Tauri stars (CTTSs) and that the field is probably dipolar at larger distances from the star, but may have a strong multipolar component close to the star. \cite{Donati07} derived the magnetic topology on the surface of CTTS V2129~Oph, and found that it has a dominant octupole component with $B_{oc} \approx 1.2 \,\mathrm{kG}$ and a weaker dipole component with $B_d \approx 0.35\, \mathrm{kG}$.
Additionally, early research on X-ray emission in neutron stars also suggested a quadrupole magnetic component \citep{Shakura91,Panchenko94}.
Thus, the magnetic field configuration is likely to differ among stars, based on the significance of the dipole component.
\cite{Lipunov78} suggested a framework to extend the spherical accretion model to the
case of stellar magnetic fields having higher-order multipole moments. \cite{vonRekowski06} performed axisymmetric simulations considering the disk-magnetosphere interaction when the magnetic field is generated by the dynamo effect and found a time-varying stellar magnetic field with a complex multipolar configuration.

Although observational and theoretical evidence suggests that stellar magnetic configurations may not be purely dipolar, most 
simulations of disk-magnetosphere interactions assume a~dipolar field configuration. 
Exceptions include, for example, \cite{LongRom07,LongRom08} and \cite{Ciec22} in non-relativistic simulations, or \cite{Das22} in the general relativistic simulations. \cite{Cikin23} studied the interaction between the disk and a neutron star with a quadrupole field configuration, concluding that the effect of the quadrupole depends on the location of the inner radius of the disk and the strength of the quadrupole field.

In this work, we examine the magnetospheric accretion through 2.5-D MHD simulations, exploring various initial magnetic configurations. These include dipolar, quadrupolar, octupolar fields, as well as combined
configurations of dipolar-quadrupolar and dipolar-octupolar fields, all within the context of disk accretion onto a rotating star. We consider the observational parameters for the star IRAS-17449+2320  characterized by a strong magnetic field ($\approx 6.2\, \mathrm{kG}$) and low rotational velocity \citep{Korcakova22}. This particular object has raised the possibility of post-mergers ocurring among FS~CMa stars, a subgroup of B[e] stars whose distinctive characteristic in the spectra is the existence of forbidden emission lines and a
strong IR excess. 

In our earlier work, we performed 2.5-D MHD simulations of disk accretion onto stars with pure dipole fields \citep{Moranchel-Basurto23,Moranchel24}, revealing distinct features in the funnel streams and associated gaps compared to cases involving a pure dipole field and magnetic diffusivity. In scenarios with significant resistivity, accretion always occurs along the midplane and produces magnetospheric ejections from the disk and stellar surface.

The paper is laid out as follows. In Section \ref{sec:model}, we describe our numerical model and the different magnetic configurations considered. The results of the simulations are given in Section \ref{sec:results}. In Section \ref{sec:discussion}, we give a brief discussion. Finally, the conclusions are given in Section \ref{sec:conclusions}.

\section{Model setup and magnetic field configurations}
\label{sec:model}

We model the evolution of the gas in a thin accretion disk and the magnetosphere region surrounding an FS~CMa type star. 
The simulations solve the MHD equations (eq. \ref{eq:momentum} to eq. \ref{eq:induction}), including both resistivity and viscosity:
\begin{equation}
    \frac{\partial \rho}{\partial t}+\vec{\nabla} \cdot (\rho \vec{v)} =0,
    \label{eq:momentum}
\end{equation}
\begin{equation}
    \frac{\partial \rho \vec{v}}{\partial t} + \vec{\nabla} \cdot \left(\rho \vec{v} \vec{v}^T -\frac{\vec{B} \vec{B}^T}{4\pi}\right) + \vec{\nabla} \left(P+\frac{B^2}{8\pi}\right) + \rho \vec{\nabla}\Phi =0,
\end{equation}

\begin{align}
    \frac{\partial e}{\partial t}& + \vec{\nabla} \cdot \left[ \vec{v} \left(e+P+ \frac{B^2}{8\pi}\right)-\frac{1}{4\pi}(\vec{v}\cdot \vec{B})\vec{B}+ (\eta \cdot \vec{J})\times \vec{B} \right]  \nonumber \\ 
    &
    =-\rho(\vec{\nabla}\Phi) \cdot \vec{v},
\end{align}
and the induction equation
\begin{equation}
    \frac{\partial \vec{B}} {\partial t}- \vec{\nabla}\times (\vec{v} \times \vec{B} - \eta \cdot \vec{J})=0,
    \label{eq:induction}
\end{equation}
\noindent where $P$ is the thermal gas pressure, $\rho$ the gas density, $\vec{v}$ the gas velocity vector,  $\vec{B}$ the magnetic flux density vector, $\vec{J}$ the electric current density vector, $\eta$ the electrical resistivity, 
 which is related with the magnetic diffusivity defined as $\nu =\eta/4\pi$, $\Phi = -GM_\star/R$ the gravitational potential, and $e$ represents the total energy density given by $e=P/(\gamma -1)+\rho v^{2}/2+ B^{2}/(8\pi)$, where $\gamma=5/3$ is the plasma polytropic index. The viscous stress tensor $\tau$ is defined as:
\begin{equation}
   \tau = \eta \left[(\nabla v)+ (\nabla v)^T -\frac{2}{3}(\nabla v)I\right]
\end{equation}  
where $\mathbf{I}$ is the unit tensor.

We employ a two-dimensional computational domain $(R,\theta)$ with three-dimensional vector fields  in a spherical coordinate system $(R, \theta, \phi)$, assuming axisymmetry around the stellar rotation axis so called 2.5-D approach. Our calculations were carried using the code PLUTO\footnote{A versatile code that offers modules for hydrodynamic, magnetohydrodynamic, relativistic physics, among other capabilities and is based on the Godunov-type numerical scheme (see https://plutocode.ph.unito.it/).} \citep{M2007}, using the HLL Riemann solver and the second-order Runge–Kutta integration in time to advance conserved variables. To compute viscous and resistive terms a second-order finite difference approximation, integrated explicitly in time, was adopted.

In Table \ref{tab:parametros} we present the values of parameters used in this work, and is divided into two sections. The first section presents the physical parameters adopted in the simulations for each component involved (the central star, the disk, and the corona). Since our study focuses on the analysis of the surroundings of FS~CMa stars, the values considered for the central star are mainly based on the observational parameters for the object IRAS-17449+2320 reported by \cite{Korcakova22}. The second section of Table \ref{tab:parametros} provides a summary of the computational grid and the used normalization values, as the simulations are performed using dimensionless equations.\footnote{Note that the evolution time of our models is less than $t=10T_0$, because the magnetic field strength in the different configurations quickly shapes the gas flow, which is the main objective of this study. In addition, for a longer evolution, the computational cost is notably increased due to the formation of very low-density regions in the magnetosphere caused by the intensity of the magnetic field itself. This could be fixed by refilling the density at each timestep. However, we prefer to avoid generating artificial density substructures in our models.
}

\subsection{Model description}
\label{model_desciption}
 
\begin{table}
\tiny
    \centering
    \begin{tabular}{|p{0.005cm}| p{2.3cm} p{2.1cm} p{2.3cm}|}
   \hline
\centering
&  & Parameter & Value \tabularnewline \cline{1-4} \\
\centering
 \parbox[t]{2mm}{\multirow{13}{*}{\rotatebox[origin=c]{90}{Physical setup}}} & Central star &  & \tabularnewline \cline{2-4} \\
  &  Mass      & $M_\star$ & $6M_\odot$  \\
  &  Radius      & $R_\star$ & $3R_\odot$  \\
  &  Magnetic field     & $B_\star$ & $6.2$ kG  \\
  &  Break-up speed &  $\Omega_{\rm br} = \sqrt{GM_\star/R_\star^3}$ & $2.966\times10^{-4}\mathrm{s}^{-1}$ \\
   & Stellar rotating rate & $\Omega_\star$ & $0.1 \Omega_{\rm br}$ \\
\tabularnewline \cline{2-4} 
\centering
 &Disk &  &  \tabularnewline \cline{2-4}   \\
  &Initial density       &  $\rho_{d0}$ & $1\times 10^{-13}$g cm$^{-3}$   \\  
  & Aspect ratio & $h$ &  $0.1$ \\
 &Viscosity      & $\alpha_\nu$ & $1.0\mathrm{R_0^2\Omega_\star}$ \\
 & Magnetic resistivity        & $\alpha_m$ & $1.0\mathrm{R_0^2\Omega_\star}$ \\
   \tabularnewline \cline{2-4} 
\centering
 &Corona &  &  \tabularnewline \cline{2-4}   \\ 
 & Initial density       &  $\rho_{\rm atm}^0$& $1\times 10^{-15}$g cm$^{-3}$  \\  
 &Viscosity      & $\alpha_\nu$ & $0.0$ \\
 & Magnetic resistivity        & $\alpha_m$ & $0.0$ \\
\tabularnewline \hline
\centering
  \parbox[t]{2mm}{\multirow{11}{*}{\rotatebox[origin=c]{90}{Computational setup}}}& Mesh &  &  \tabularnewline \cline{2-4}  \\
&Radial extension         & $R$ &  $[1R_\star , 60R_\star]$ \\
&Colatitude extension         & $\theta$ &  $[0 , \pi]$ \\
&Radial resolution         & $N_R$ &  436 cells \\
&Colatitude resolution         & $N_\theta$ &  200 cells \\
& Radial spacing        & log & logarithmic \\
\tabularnewline  \cline{2-4}
 \centering
   & Normalization  &  &  \tabularnewline \cline{2-4}  \\
& Length         & $R_0=R_{\star}$ &  $2.087\times 10^{11}$ cm \\
& Density         & $\rho_{0}=\rho_{d0}$ & $ 1 \times 10^{-13}$ g cm$^{-3}$ \\
& Velocity         & $V_{K0}=\sqrt{GM_\star/R_0}$ &  $6.193 \times 10^7$ cm/s \\
&Time         & $t_0=R_0/v_{K0}$ &  $3.37 \times 10^3$ s  \\
& Rotation period         & $T_0=2\pi t_0$ &  $2.12 \times 10^4$ s \\
&Reference field & $B_{0}=\sqrt{4\pi\rho_{d0} V_{K0}^{2}}$ & $69.42$ G\\
&Star magnetic field & $B_\star = \mu B_{0}$ & $6.2$kG
with $\mu=89.3$\\
& Accretion Rate & $\dot{M}_0={\rho_{0} V_{K0}R_0^{2}}$ & $4.257\times 10^{-9} M_\odot$ yr$^{-1}$\\
\hline
\end{tabular}
\caption{Initial conditions and parametrization of the setup used in our simulations. Note that the parameter $\mu$ appearing in $B_\star$ specifies the ratio between the stellar surface magnetic field and the reference magnetic field strength.}
\label{tab:parametros}
\end{table}

The density $\rho_{d}$ and pressure $P_{d}$ of the gas in the accretion disk are given by the following expressions \citep[e.g.,][]{Zanni2009,Moranchel24}:

\begin{equation}
    \rho_d (R,r) = \rho_{d0}\left\{\frac{2}{5h^2}\left[\frac{R_0}{R}-\left(1-\frac{5h^2}{2}\right)\frac{R_0}{r}\right]\right\}^{3/2}, 
    \label{eq:rho_}
\end{equation}
\begin{equation}
       P_d = h^2\rho_{d0}v_{K0}^2\left(\frac{\rho_d}{\rho_{d0}}\right)^{5/3}, 
\end{equation}
where $\rho_{d0}$ and $v_{K0}$ are the density and Keplerian velocity at the midplane of the disk at $R=R_0$, 
$h=C_s/v_K$ is the disk aspect ratio, with $C_s$ the isothermal sound speed and $v_{K}= \sqrt{GM_\star/r}$.
The values of these parameters are given in Table \ref{tab:parametros}.  Note that $r=R\sin{\theta}$ is the cylindrical radius.

The initial disk atmosphere is defined with the assumption
of a polytropic gas in hydrostatic equilibrium, without rotation.  
The density and pressure are given by: 
\begin{equation}
    \rho_{\mathrm{atm}} (R)=\rho_{\mathrm{atm}}^0\left(\frac{R_\star}{R}\right)^{\frac{1}{\gamma-1}},
\end{equation}
and
\begin{equation}
    P_{\mathrm{atm}}(R)=\rho_{\mathrm{atm}}^0\frac{\gamma-1}{\gamma}\frac{GM_\star}{R_\star}\left(\frac{R_\star}{R}\right)^{\frac{\gamma}{\gamma-1}},
\end{equation}
with $\gamma =5/3$ \citep[e.g.][]{Zanni2009}. The ratio of the coronal to disk density at $R=R_{0}$ and $z=0$ is $\rho_\mathrm{atm}^0/\rho_{d0}=0.01$, a value adopted in all models.

\begin{table}
    \centering
    \begin{tabular}{p{2.5cm} p{1.5cm} p{1.5cm}}
        \hline
\centering
    Model & $B_\star^d/B_\star^q$  & $B_\star^d/B_\star^{oc}$ \\
       \hline
\centering
        D & - &  - \\ 
\centering
        Q & - & - \\  
\centering
        $\mathrm{DQ}_\mathbf{17}$  & 1/7 & -\\

\centering
        $\mathrm{DQ}_\mathbf{12}$  & 1/2 & - \\
\centering
        $\mathrm{DQ}_\mathbf{1}$ & 1&  - \\

\centering
        $\mathrm{DQ}_\mathbf{3}$ & 3 & - \\
\centering
        $\mathrm{DO}_\mathbf{17}$ & -  & 1/7 \\
\centering
        $\mathrm{DO}_\mathbf{13}$ & -  & 1/3 \\
\hline
    \end{tabular}
    \caption{Summary of the ratios of the magnetic polar strength in the different configurations considered in our numerical models.} 
    \label{tab:models_b}
\end{table}

To model the star's rotation and the boundary conditions of a perfect accretor, we follow the descriptions given in previous works \citep{Zanni2009,Cemeljic2019,Moranchel24}. 

\begin{figure}
\centering
 \begin{subfigure}{1.05\textwidth}
\includegraphics[width=0.43\textwidth]{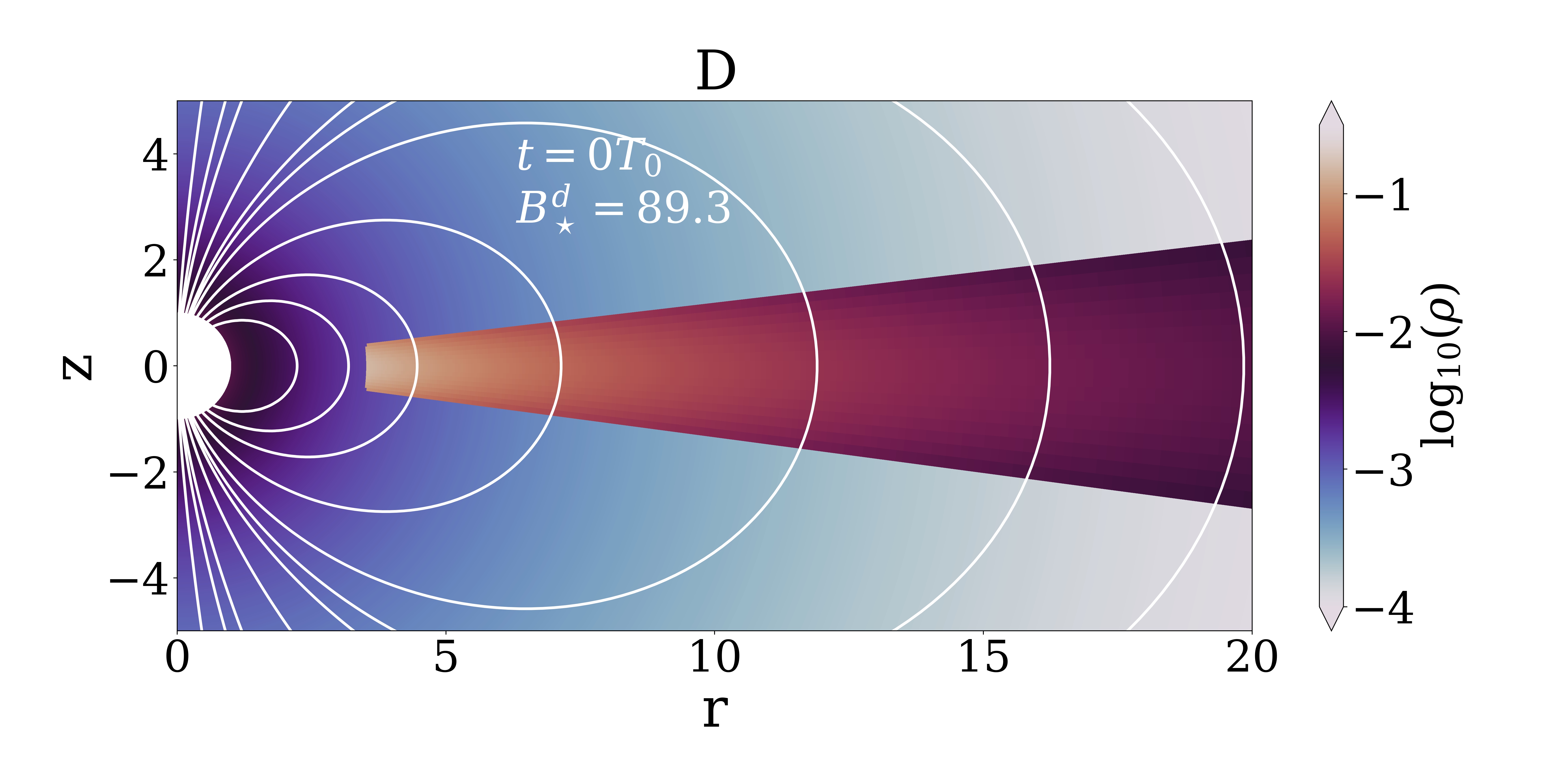}
 \end{subfigure}
 \begin{subfigure}{1.05\textwidth}
\includegraphics[width=0.43\textwidth]{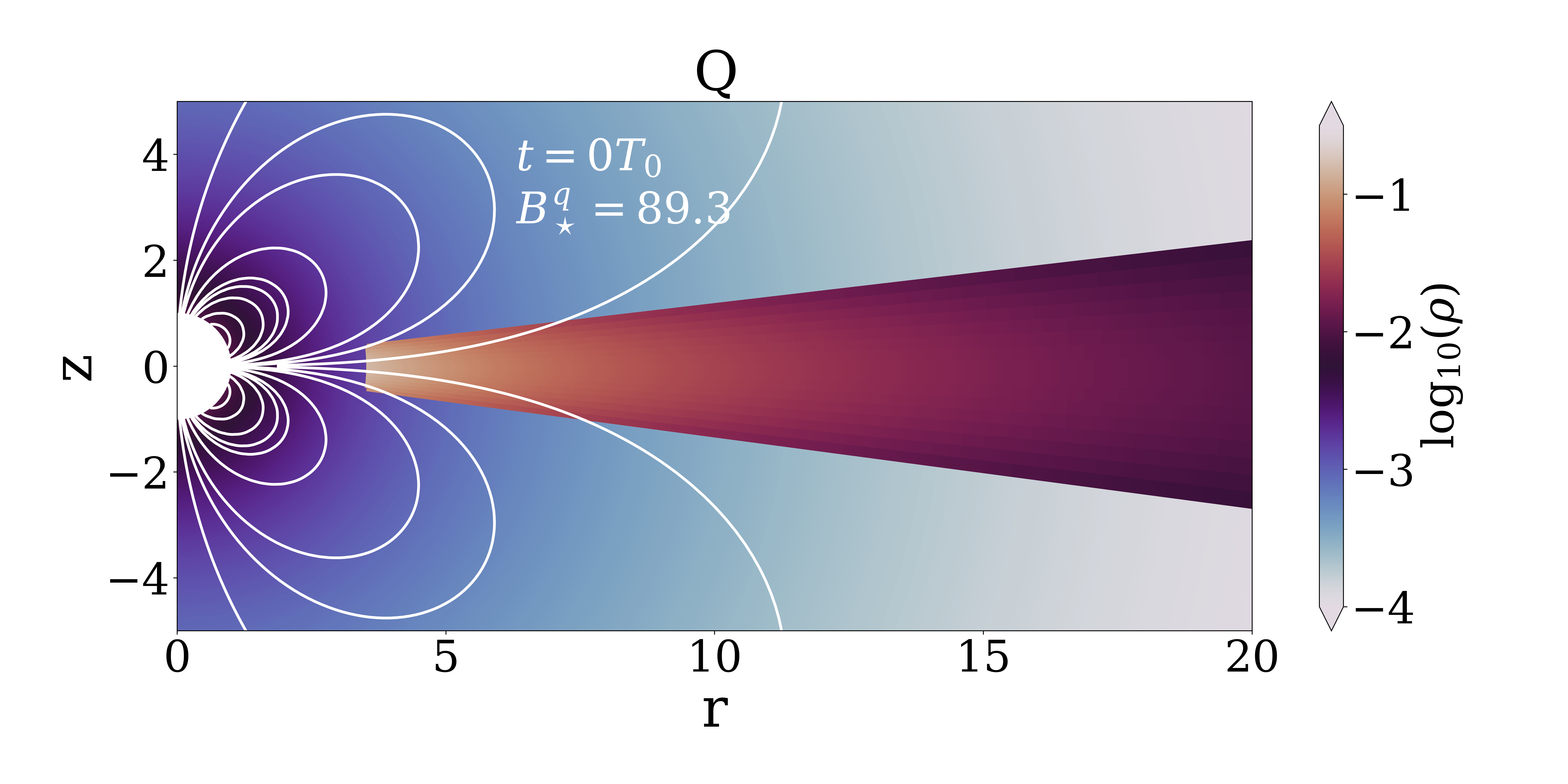}
 \end{subfigure}
 \begin{subfigure}{1.05\textwidth}
     \includegraphics[width=0.43\textwidth]{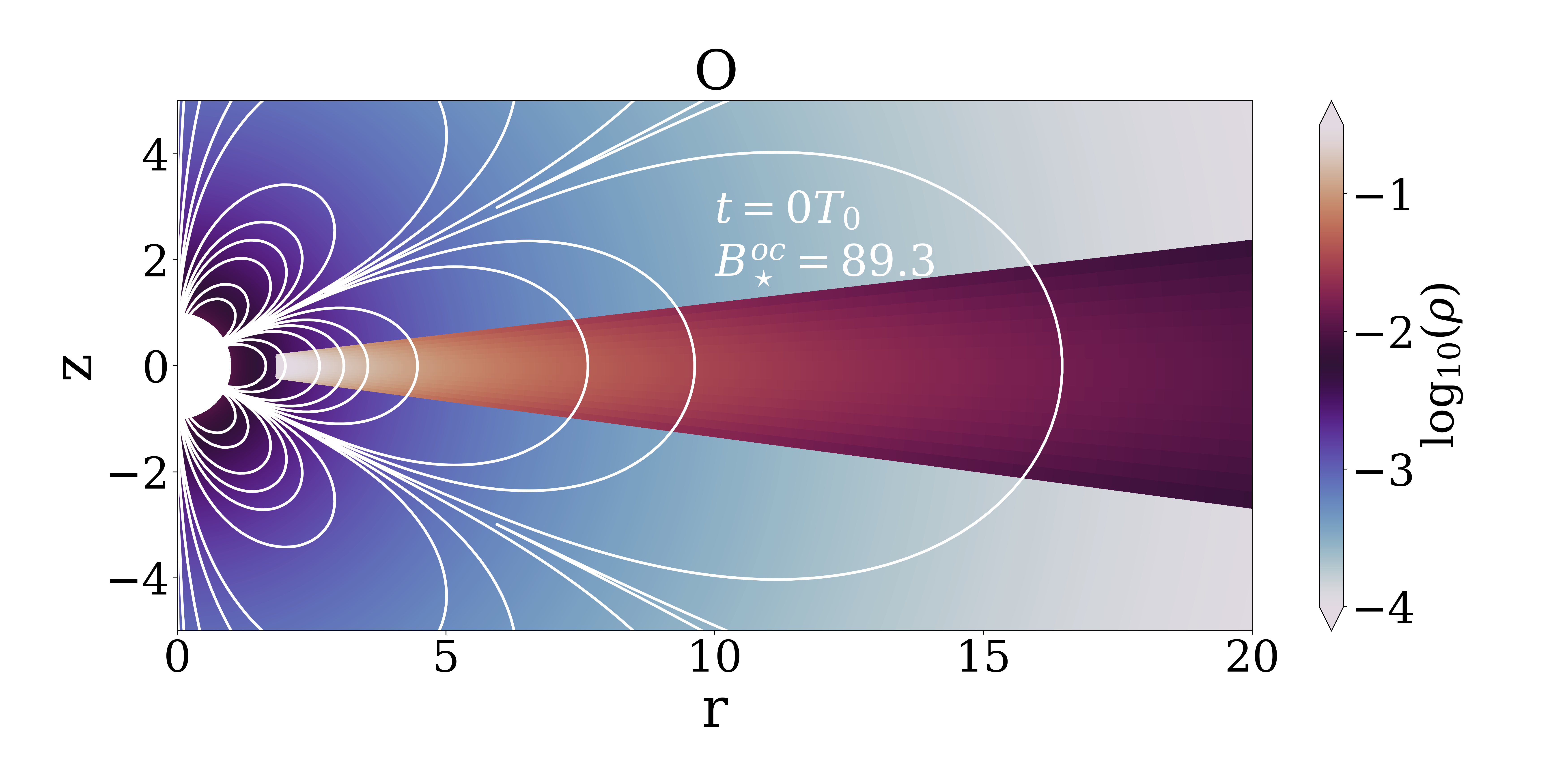}
 \end{subfigure}
 \begin{subfigure}{1.05\textwidth}
     \includegraphics[width=0.43\textwidth]{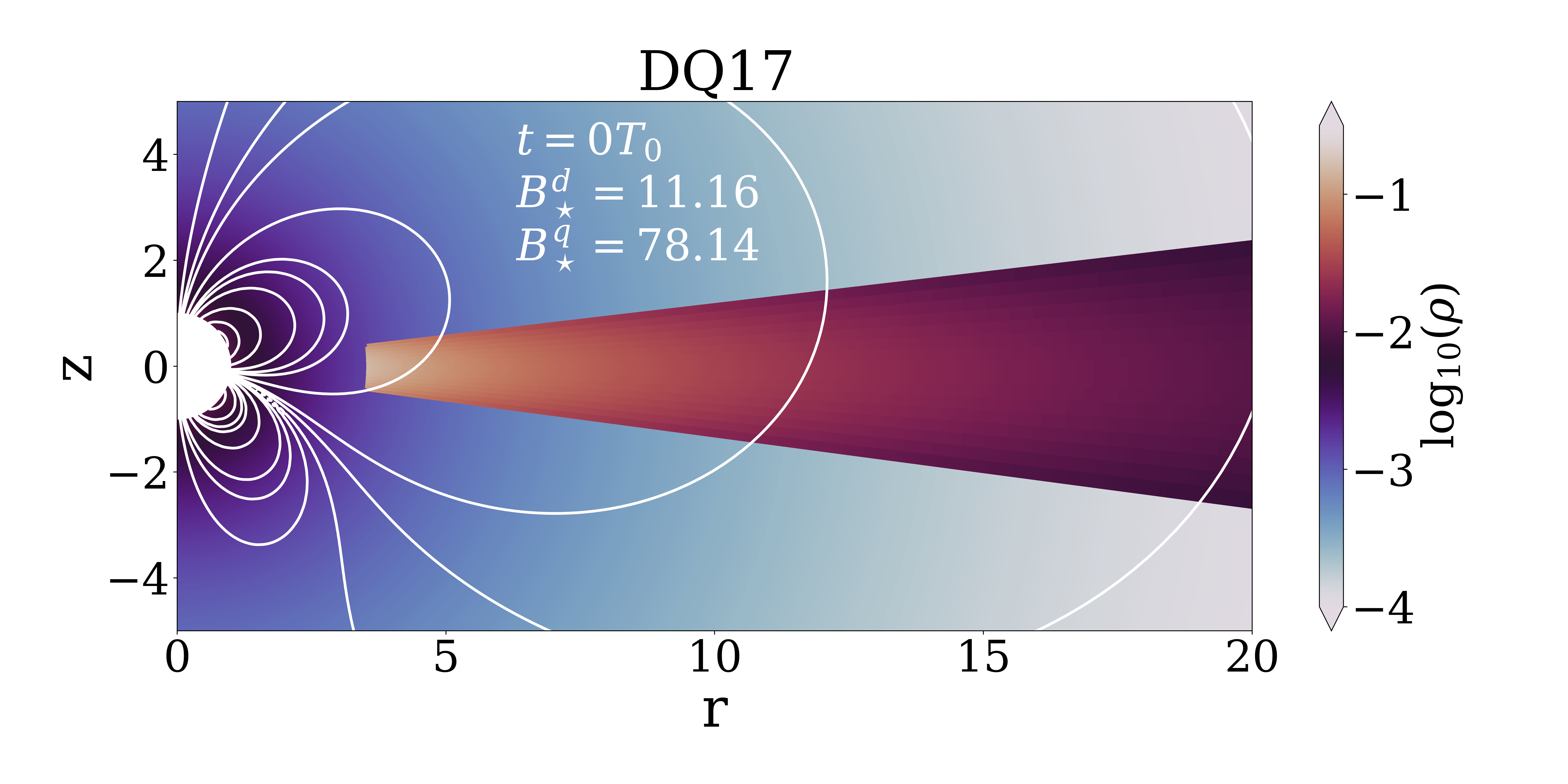}
 \end{subfigure}
 \begin{subfigure}{1.05\textwidth}
     \includegraphics[width=0.43\textwidth]{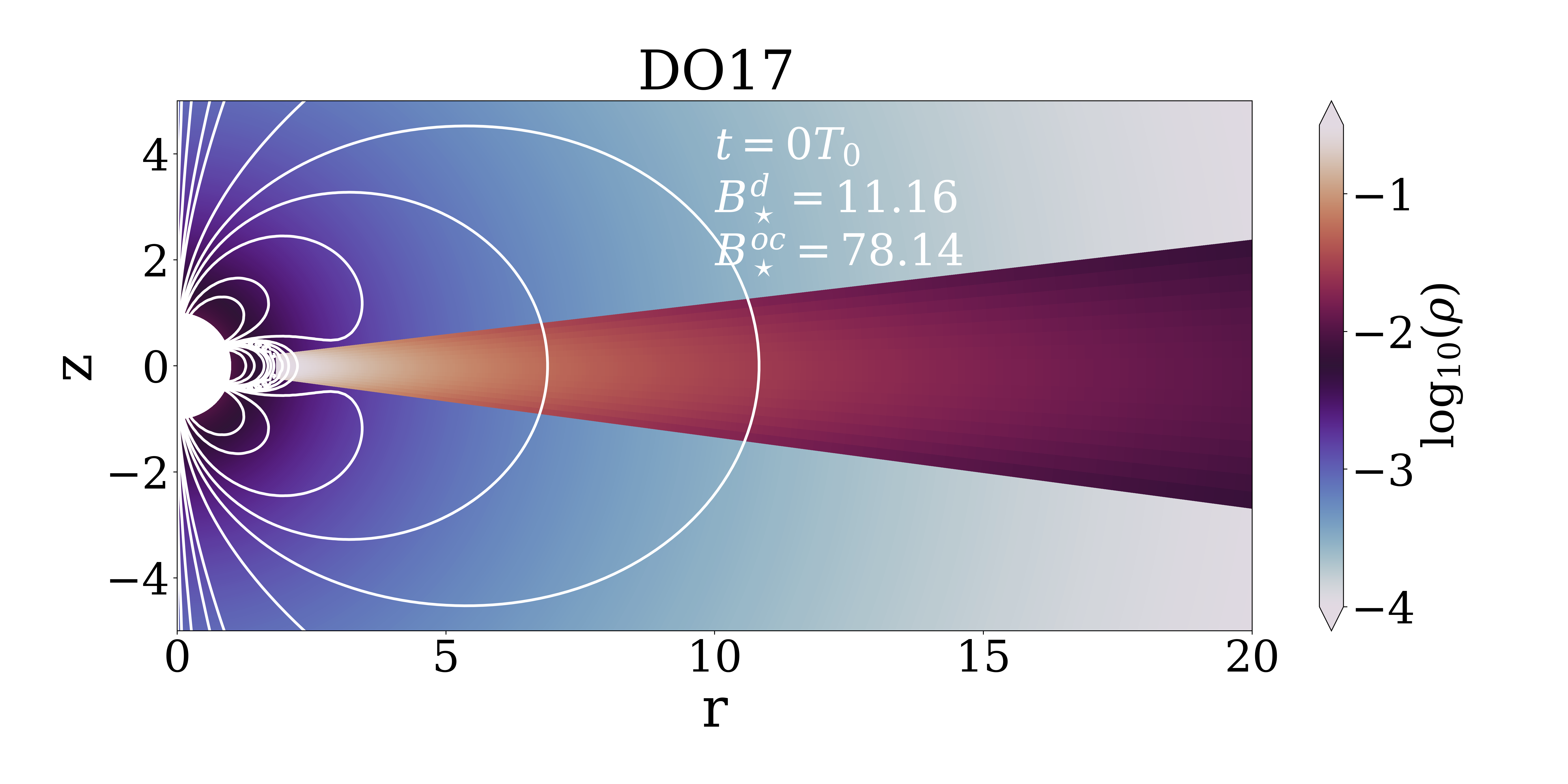}
 \end{subfigure}
 \caption{Initial density (in units of $\rho_0$) and magnetic field lines (white curves) for model D (first panel), model Q (second panel), model O (third panel), model DQ (fourth) and model DO (fifth). The letters D, Q, O, DQ and DO represent dipolar, quadrupolar, octupolar, dipolar-quadrupolar and dipolar-octupolar configurations, respectively. In all cases, the sum of the different multipolar contributions to the stellar magnetic field is equal to $89.3$. Here, $r$ and $z$ represent the cylindrical coordinates, measured in units of the stellar radius. }
 \label{fig:initial comparison}
 \end{figure}

\begin{figure*}
    \begin{subfigure}{0.5\textwidth}
      \includegraphics[scale=0.2]{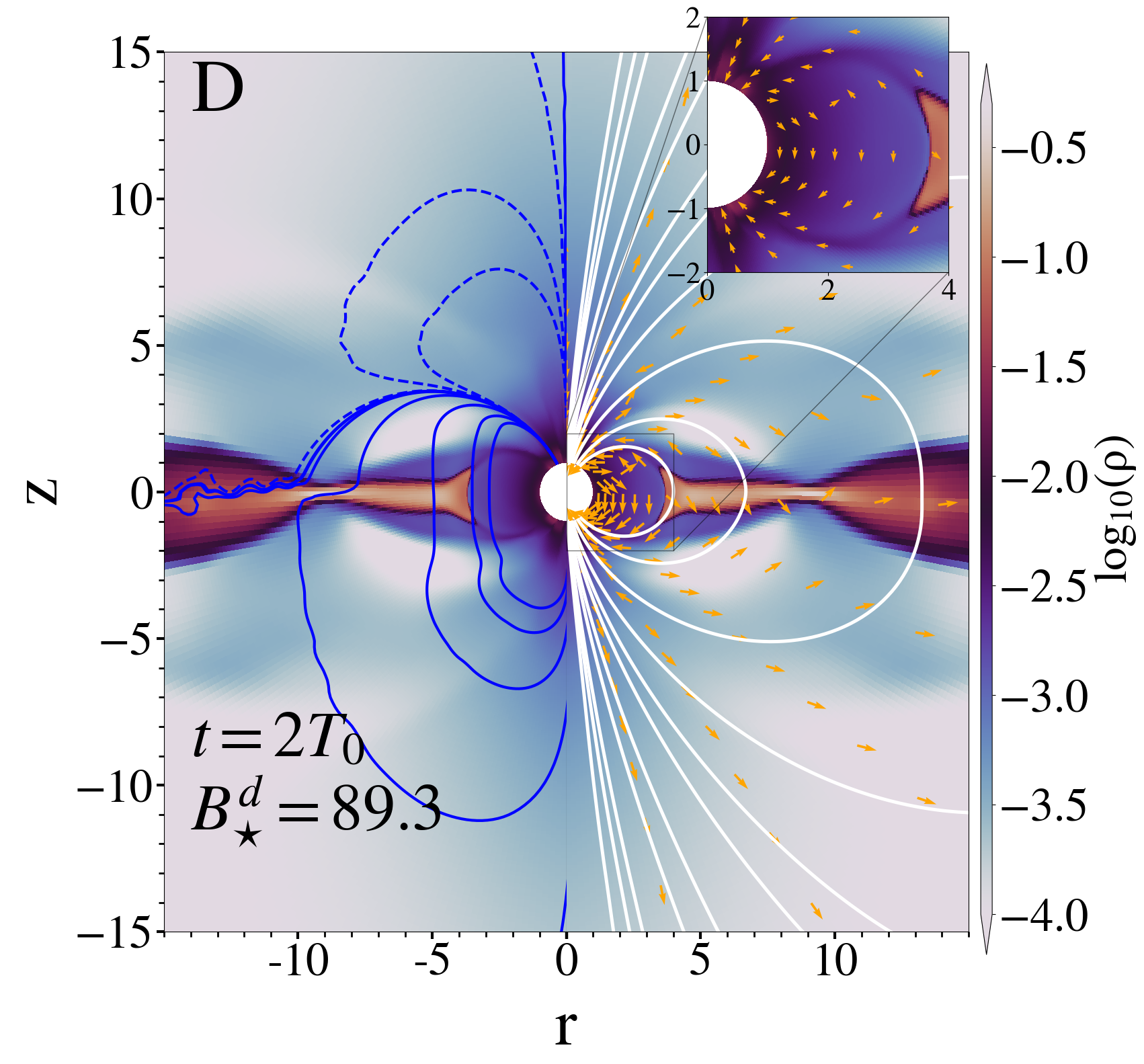}
    \end{subfigure}
    \begin{subfigure}{0.5\textwidth}
    \includegraphics[scale=0.2]{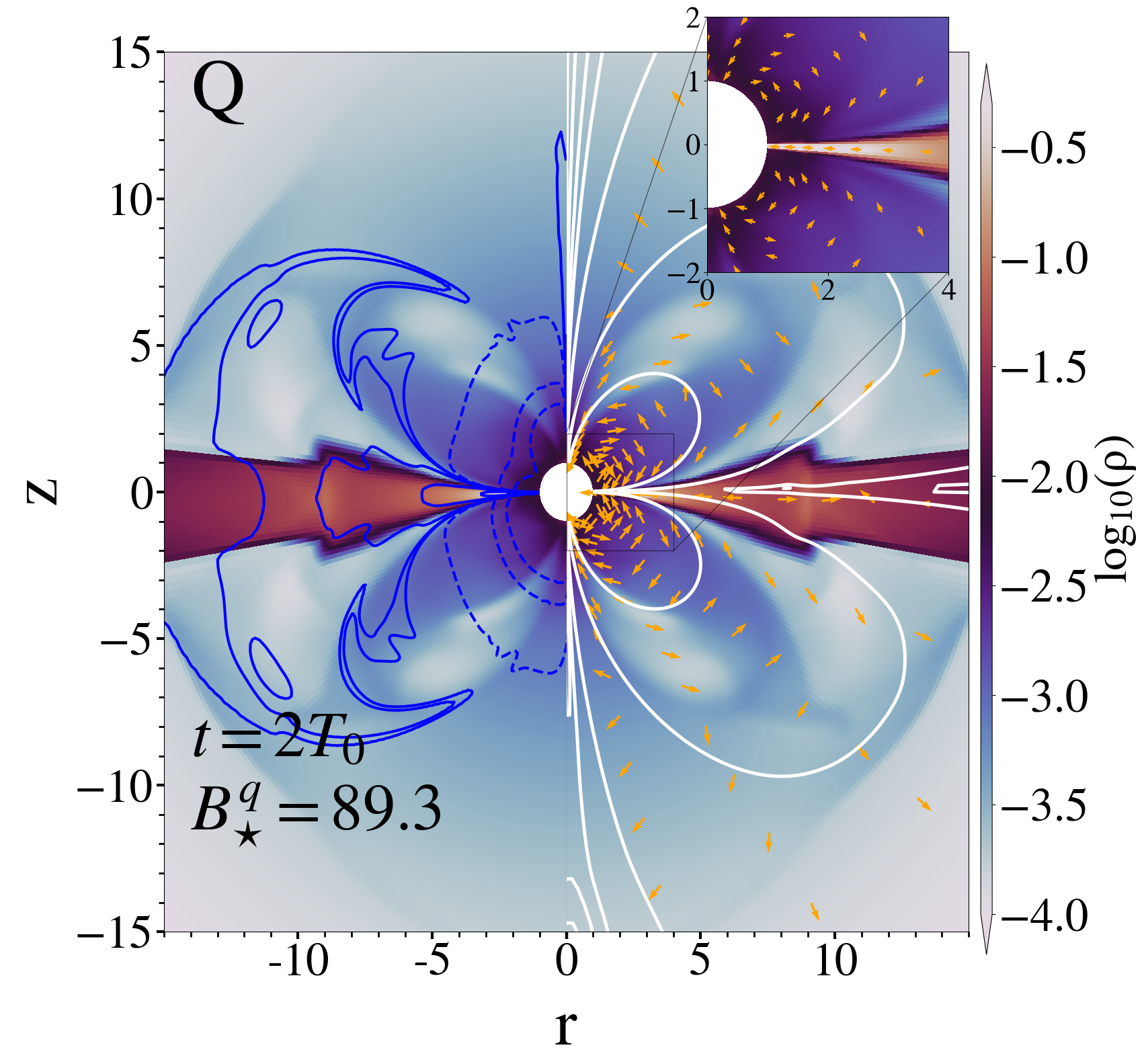}
    \end{subfigure}
   \caption{Density maps (in units of $\rho_0$) at $t=2T_0$ for a dipolar (D model) magnetic configuration (left panel) and for a quadrupolar (Q model) magnetic configuration (right panel). Blue solid and dashed lines on the left side of each panel show the angular momentum flux $\mathbf{f}_B$
   carried by the field (see Eq. \ref{eq:fb}), with solid blue indicating positive flux and dashed blue indicating negative flux, while in the right side of each figure the yellow vectors show the velocity and the white lines represent the poloidal magnetic field lines. Lastly, inserts show a zoomed region of the density map for $R<R_\mathrm{co}$ where it is clearly seen how the gas follows different paths towards the star for the D and Q models. However, it is worth noting that, in both cases, the gas flow toward the star is symmetric about the midplane.
   }
 \label{fig:ComparisonDQ}
\end{figure*}

\subsection{Magnetic configurations}

In our work, the magnetic configurations exhibit axial symmetry around the rotational axis of the star and disk. Additionally, we assume that the magnetic moments are aligned with the stellar rotational axis, allowing us to express the magnetic field components in terms of the polar magnetic field strength (see \cite{Gregory10} for a derivation) as follows. 

For the pure dipolar case:
 \begin{equation}
 \centering
     B^{d}_R (R,\theta)= B_\star^{d} \left( \frac{R_\star}{R}\right)^3 \cos{\theta},
     \label{eq:B_r_dipolo}
  \end{equation}
 and
 \begin{equation}
      B^{d}_{\theta} = \frac{1}{2} B_\star^{d} \left(\frac{R_\star}{R}\right)^3 \sin{\theta},
      \label{eq:B_teta_dip}
 \end{equation}
\noindent For a quadrupolar configuration, the components are:
 \begin{equation}
    B^{q}_R(R,\theta) = \frac{1}{2} B_\star^{q} \left( \frac{R_\star}{R}\right)^4 (3\cos^2\theta -1),
    \label{eq:B_r_cua}
 \end{equation}
 and
 \begin{equation}
      B^{q}_{\theta} =  B_\star^{q} \left( \frac{R_\star}{R}\right)^4 \sin{\theta} \cos{\theta}.
      \label{eq:B_teta_cua}
 \end{equation}
 Finally, for an octupolar field configuration:
 \begin{equation}
     B^{oc}_R (R, \theta)= \frac{1}{2} B_\star^{oc} \left( \frac{R_\star}{R}\right)^5 (5\cos^3{\theta}-3\cos\theta),
     \label{eq:B_r_oct}
 \end{equation}
and
 \begin{equation}
     B^{oc}_\theta (R,\theta)= \frac{3}{8} B_\star^{oc} \left( \frac{R_\star}{R}\right)^5 (5\cos^2\theta \sin\theta - \sin\theta).
     \label{eq:B_teta_oct}
 \end{equation}
 
 In Equations (\ref{eq:B_r_dipolo}) - (\ref{eq:B_teta_oct}), $B_\star^{d}$, $B_\star^{q}$ and $B_\star^{oc}$ are the polar magnetic field strengths of the dipole, quadrupole, and octupole at the stellar rotation pole, respectively. 

Additionally, we explore models that combine a dipole with a quadrupole configuration, as well as a dipole with an octupole configuration. More specifically, the magnetic field is given by $\mathbf{B}^{dq} = \mathbf{B}^{d} + \mathbf{B}^{q}$ for dipole plus quadrupole, and $\mathbf{B}^{do}= \mathbf{B}^{d} + \mathbf{B}^{oc}$, for the dipole plus octupole configuration.

A compilation of the potential vectors $\mathbf{A}$ and the magnetic flux functions $\psi$ for
the dipolar, quadrupolar and octupolar configurations is given in Appendix \ref{app:Apluspsi}.

To specify the total magnetic field strength on the stellar surface,
we use a dimensionless parameter $\mu$, defined as the
ratio of the measured magnetic field strength on the stellar 
surface $B_{\star}$ to the reference field strength 
$B_{0}\equiv \sqrt{4\pi \rho_{d0} V_{K0}^{2}}$. For $B_{\star}=6.2$ kG and for the values of $\rho_{d0}$ and $V_{K0}$ adopted in
this work (see Table \ref{tab:parametros}), $B_{0}=69.42$ G and therefore $\mu = B_{\star}/B_0=89.3$.


\subsection{Diagnostics}
\subsubsection{Angular momentum flux}

To analyze the different behaviors in the mass flow accreted by the star through this study, we consider 
the angular momentum fluxes carried by the magnetic field $\mathbf{f}_B$, and by the matter
$\mathbf{f}_m$, as given by
\begin{equation}
\mathbf{f}_B=\frac{rB_\phi\mathbf{B}_p}{4\pi}
    \label{eq:fb}
\end{equation}
and
\begin{equation}
\mathbf{f}_m=-\rho r v_\phi\mathbf{v}_p,
    \label{eq:fm}
\end{equation}
respectively, with $\mathbf{B_p}$ and $\mathbf{v_p}$ the poloidal magnetic field and poloidal velocity vectors.

\subsubsection{Disk truncation radius}
\label{subsec:rt}

The most commonly adopted definition of the truncation radius $R_{\rm trunc}$ is inferred from the condition:
$B^{2}/(8\pi) = \rho_{d} v_{K}^{2}$, with $B$ and $\rho_{d}$ evaluated at the midplane of the disk (e.g., \cite{Hartmann16}; \cite{zhu24}). If the magnetic field
is a dipole:
\begin{equation} 
\frac{R_{\rm trunc,th}}{R_{\star}} = \left(\frac{(B_{\star}^{d})^{2} R_{\star}}{32\pi GM_{\star}\rho_{d0}}\right)^{2/7},
\label{ec:rt_dip}
\end{equation}
whereas for a pure octupole magnetic field:
\begin{equation} 
    \frac{R_{\rm trunc,th}}{R_{\star}}= \left(\frac{9(B_{\star}^{oc})^{2} R_{\star}}{2^{8}\pi GM_{\star}\rho_{d0}}\right)^{2/15}.
    \label{ec:rt_oct}
\end{equation}
Using the values provided in Table \ref{tab:parametros}, if the magnetic field is dipolar with a
strength of 
$B_{\star}^{d}=6.2$ kG, the truncation radius is $R_{\rm trunc}=3.48R_{\star}$. Assuming instead an
octupolar magnetic field of the same strength, we obtain $R_{\rm trunc}=1.82R_{\star}$.

We emphasize that the initial inner radius of the disk, $R_\mathrm{d}$, for the models with configurations dipolar, quadrupolar and the compositions of dipolar plus quadrupolar field were fixed initially at $R=3.48R_\star$, which is the value obtained for the truncation radius considering a purely dipolar magnetic configuration using Eq. (\ref{ec:rt_dip}) and the values in Table \ref{tab:parametros} . It is also  equivalent to $\sim 0.75$ times the corotation radius, $R_\mathrm{co}=(GM_\star/\Omega_\star^2)^\frac{1}{3}=4.65$, calculated from the same
Table \ref{tab:parametros} . On the other hand, for the models with dipolar plus octupolar configuration we consider that the inner radius of the disk is located at $R=1.82R_\star$.

\section{Results}
 \label{sec:results}
In this section we present the results of the different models labeled in Table \ref{tab:models_b}. Here, the letters D, Q, O, DQ and DO represent dipolar, quadrupolar, octupolar, dipolar-quadrupolar and dipolar-octupolar configurations, respectively. Figure \ref{fig:initial comparison} shows the initial gas density (in units of $\rho_0$), and the initial magnetic flux lines (white lines) for five different magnetic strength ratios. This figure shows each of the different shapes followed by the magnetic field lines, based on the contribution of each magnetic moment.

\subsection{Pure dipole and pure quadrupole configurations}

We start our study with the pure dipole case with dimensionless $B_\star^{d} = 89.3$ (recall that $R_\star=1$). The left panel in Fig. \ref{fig:ComparisonDQ} shows the density and the streamlines of angular momentum flux (solid and dashed blue lines) carried by the magnetic field lines at a time of $2T_0$, with $T_0$ the orbital period
at $R_{\star}$. 
The matter accretes through two funnel streams, forming arc-like structures near the magnetic poles. 
The formation of the funnels occurs at the truncation radius $R_\mathrm{trunc}\simeq3.48R_\star$ predicted by Eq. (\ref{ec:rt_dip}), and no signs of accretion are observed in the disk's midplane.
Additionally, the disk undergoes vertical compression from $R = R_\mathrm{trunc}$ to 
$R \simeq 10R_\star$, whereas it undergoes a slight thickening beyond $10R_\star$.

The right panel of Fig. \ref{fig:ComparisonDQ} shows the case of a quadrupolar magnetic configuration which is slightly more complex than the previous one.
In this configuration, the streamlines of angular momentum flux carried by the magnetic field lines form mirror lobes relative to the disk's midplane (see the second panel in Fig. \ref{fig:initial comparison}). As a result, the angular momentum flux carried by the magnetic field lines converges toward the star, very close to the disk's midplane. This causes a strong compression of the disk gas for $R\leq R_\mathrm{co}$ and expansion for $R_\mathrm{co}<R\leq10R_\star$. Therefore, at $2T_0$, we find a cone-shaped structure near the star that produces gas accretion only in the midplane of the disk. Furthermore, in the coronal region, we find that this magnetic field configuration
produces the formation of prolate-spheroid ejections slightly inclined with respect to the midplane.
 
In both models (D and Q), at $t=2T_0$, the magnetic field lines (solid white lines) are modified, and as a consequence, the magnetic field alters the equilibrium state. However, the forces are not strong enough to disrupt the disk.

On the other hand, from Fig. \ref{fig:ComparisonDQ} it can be seen that the shape and behavior of the magnetic field lines differ significantly between the dipolar and quadrupolar models.
For instance, in the dipolar model, a closed magnetosphere forms near the star with a set of inflated field lines at larger distances.
The field lines are observed to inflate and enter through the poles of the star, producing a slight asymmetry in the region of the flattened disk. In contrast, in the quadrupolar case, the poloidal lines enter through the midplane preserving a symmetry in the disk. Finally, we note that in our simulations, inflation of the poloidal lines in the quadrupolar configuration seems somewhat suppressed which could be explained by a relatively dense corona.

\begin{figure*}[htbp]
  \begin{subfigure}{0.175\textwidth}
      \includegraphics[scale=.19]{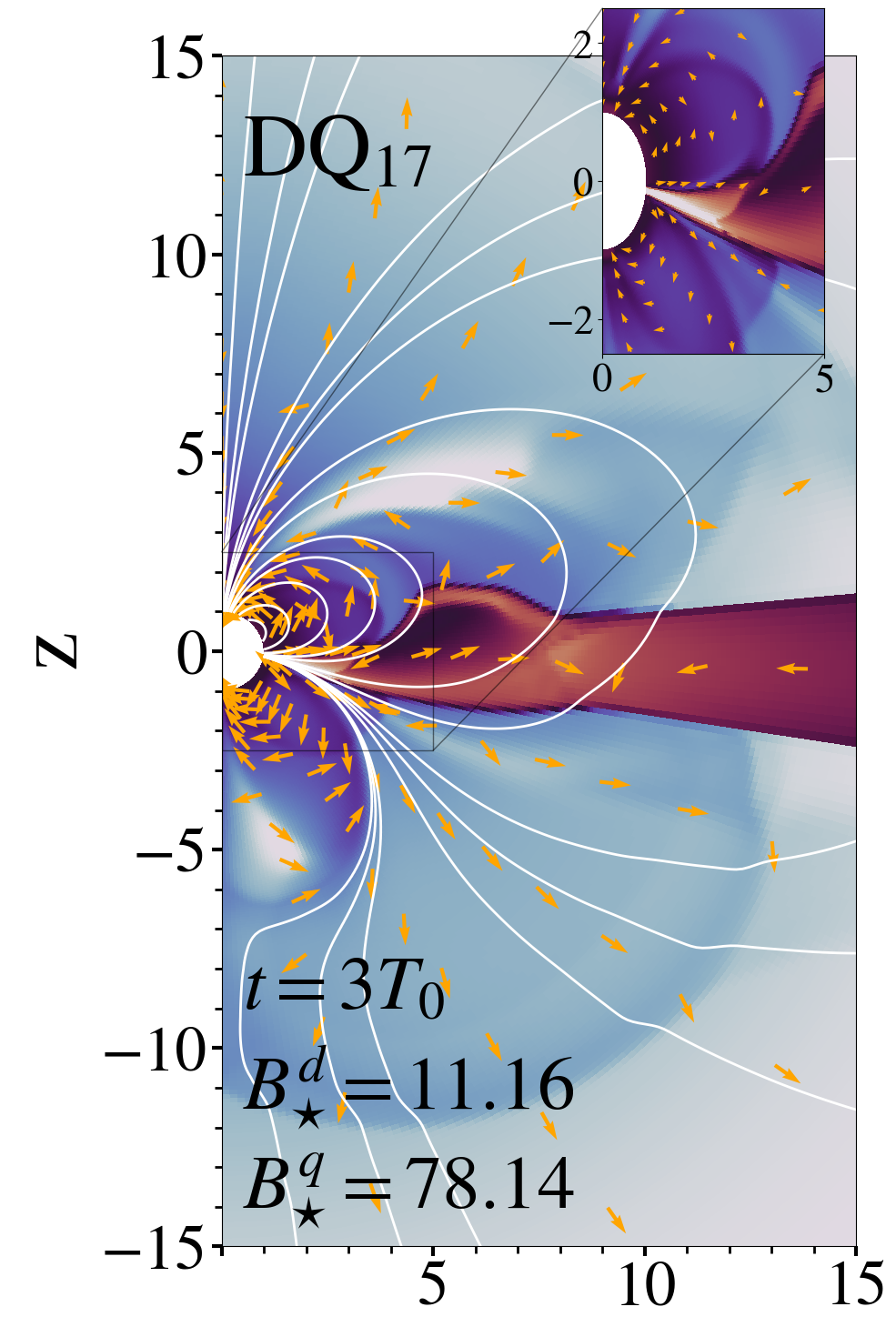}
    
   \end{subfigure}
   \hfil
   \begin{subfigure}{0.15\textwidth}
       \includegraphics[scale=0.19]{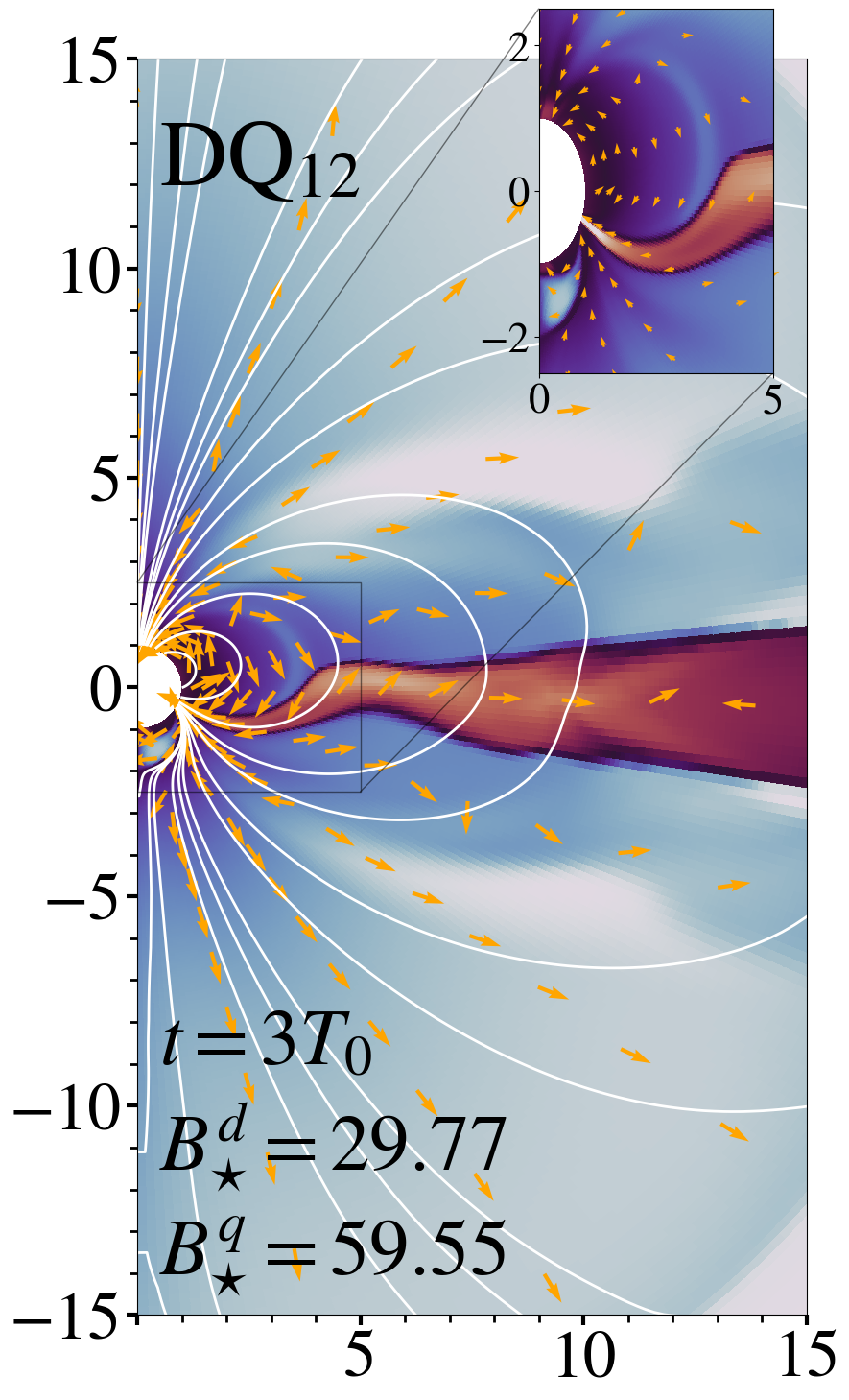}
     
   \end{subfigure}
   \hfil
   \begin{subfigure}{0.15\textwidth}
      \includegraphics[scale=0.19]{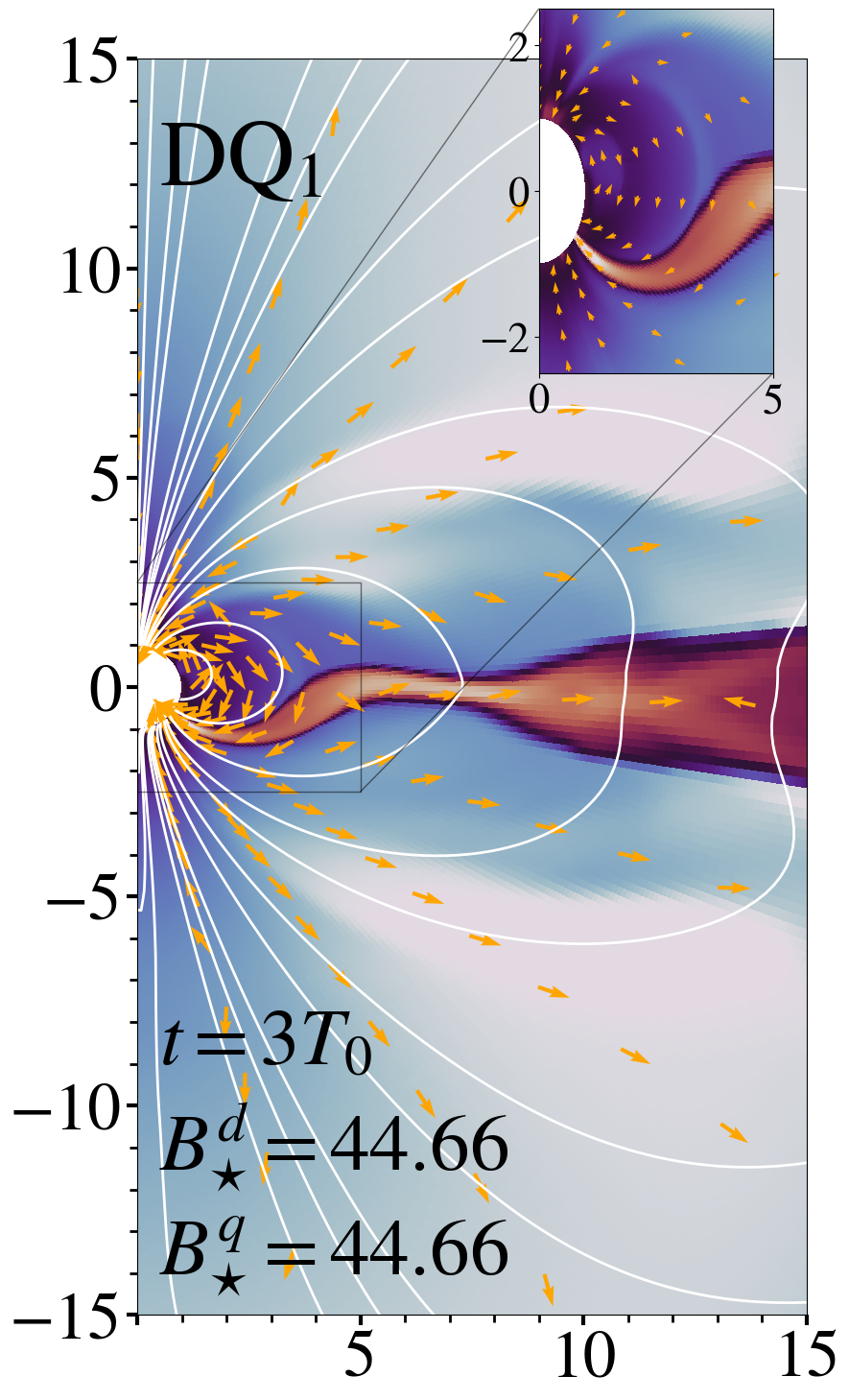}
    
    \end{subfigure}
    \hfil
    \begin{subfigure}{0.15\textwidth}
    \includegraphics[scale=0.19]{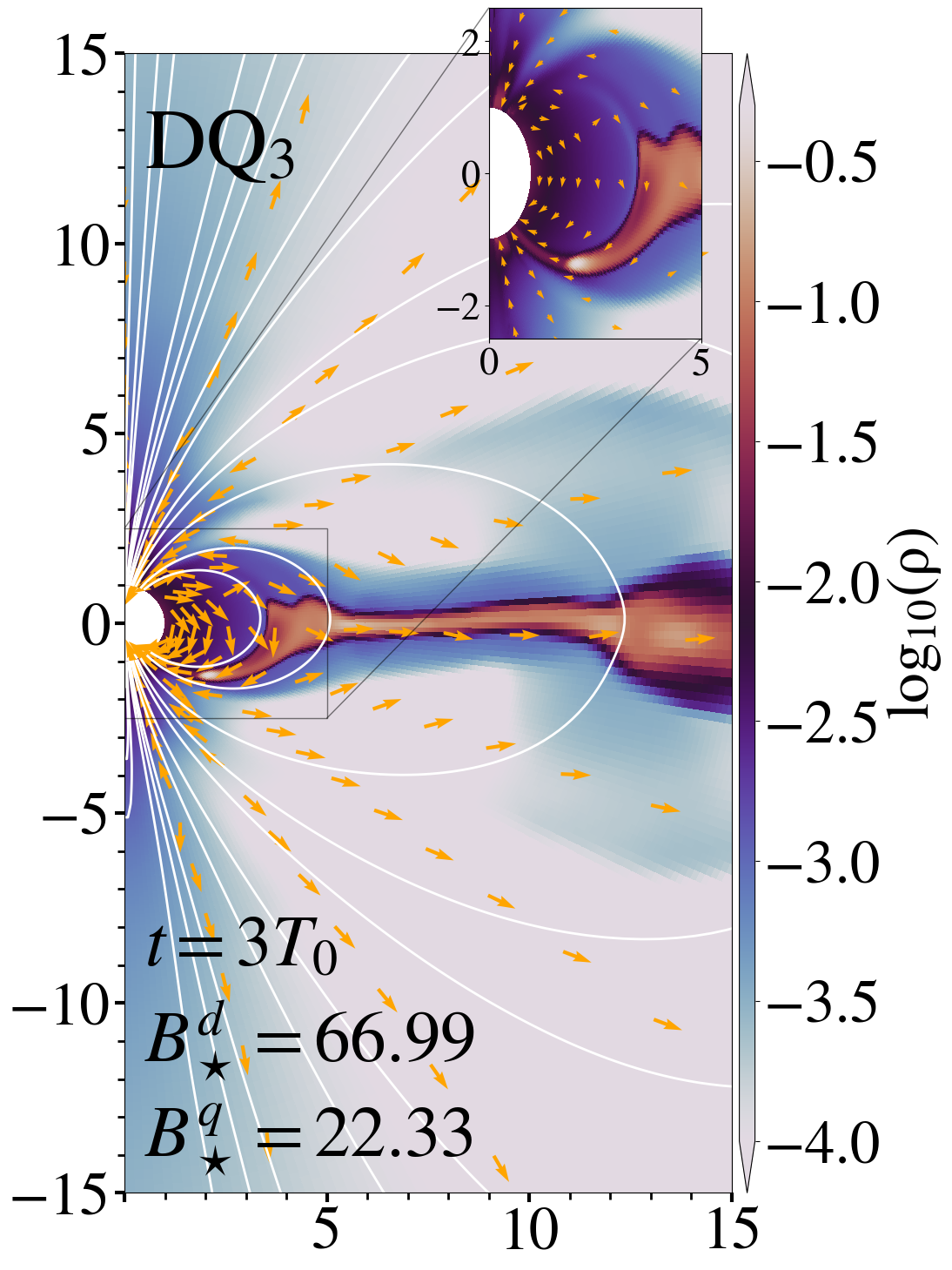}
     
    \end{subfigure}
 \caption{Gas density maps for DQ models (see Table \ref{tab:models_b}), at $t=3T_0$. We overlapped the yellow vectors to show the direction of the gas velocity and the poloidal magnetic field lines in each case (solid white lines). Note that the gas follows the magnetic field lines forming only a twisted funnel flow that enters the star below the midplane of the disk in all cases. Here the inserts show a zoom region of the density maps contained within the corotation radius.
  }
 \label{fig:dq_pollines}
 \end{figure*}
\subsection{Dipolar-quadrupolar combinations}

To model the corona of FS~CMa stars, we consider various configurations formed by combining dipolar and quadrupolar fields, as indicated in Table \ref{tab:models_b} with the letters DQ. The subscript under the logical name "DQ" represents the ratio $B_\star^d/B_\star^q$.

Figure \ref{fig:dq_pollines} shows the models $\mathrm{DQ}_\mathbf{17}$, $\mathrm{DQ}_\mathbf{12}$, $\mathrm{DQ}_\mathbf{1}$ and $\mathrm{DQ}_\mathbf{3}$, at $t=3T_{0}$. An inspection of this figure 
reveals that the last three models exhibit a flattened region of the disk that extends from the truncation radius to approximately $R \simeq 10R_\star$,
as the dipole strength increases.
Furthermore, accretion occurs mainly through the formation of a funnel flow below the midplane of the disk ($\theta>\pi/2$). When the quadrupole polar strength dominates, the flattened region of the disk is considerably reduced and twisted, resembling an extension of the funnel flow. Remarkably, the second funnel flow does not form above the midplane, because the magnetic field lines are more inflated and enter the star closer to the north pole (see the solid white lines in Figure \ref{fig:dq_pollines} for $z>0$).

In the particular model $\mathrm{DQ}_\mathbf{17}$, we find that, as the quadrupole magnetic polar strength dominates over the dipole, the gas tends to follow the magnetic field lines closer to the disk's midplane. However, the effect of the dipole is still visible, as a twisted funnel flow forms for $R < 10R_\star$ above the disk's midplane, leading to an inflated region at the edge of the disk. Note that although this funnel forms above the disk, it enters the star below the midplane (see insert in the first panel of Fig. \ref{fig:dq_pollines}). In fact, in this model we can see the formation of two gas lobes, one above and one more elongated below the midplane. Additionally, unlike in the pure dipole case, here the gas flow in the funnel can vary because the path traced by the field lines changes due to their expansion, resulting in variable accretion of matter (see Subsection \ref{subsec:mass_accreted}). In other words, in this dipole-quadrupole combination, intermittent gas accretion can occur; this latter accretion could be compared with observations.
Finally, the substructures that form in the corona region, reminiscent of wavy magnetospheric ejections, also become more asymmetric as the quadrupole polar strength dominates.

\subsection{Complex magnetic configuration}

We have conducted simulations using a dipolar plus octupolar magnetic field, labeled as $\mathrm{DO}_{\textbf{17}}$ and $\mathrm{DO}_{\textbf{13}}$ in Table \ref{tab:models_b} . Let us first analyze the case $\mathrm{DO}_{\textbf{13}}$. 
 
\begin{figure*}
    \centering
\begin{subfigure}{0.48\textwidth}
      \includegraphics[scale=.20]{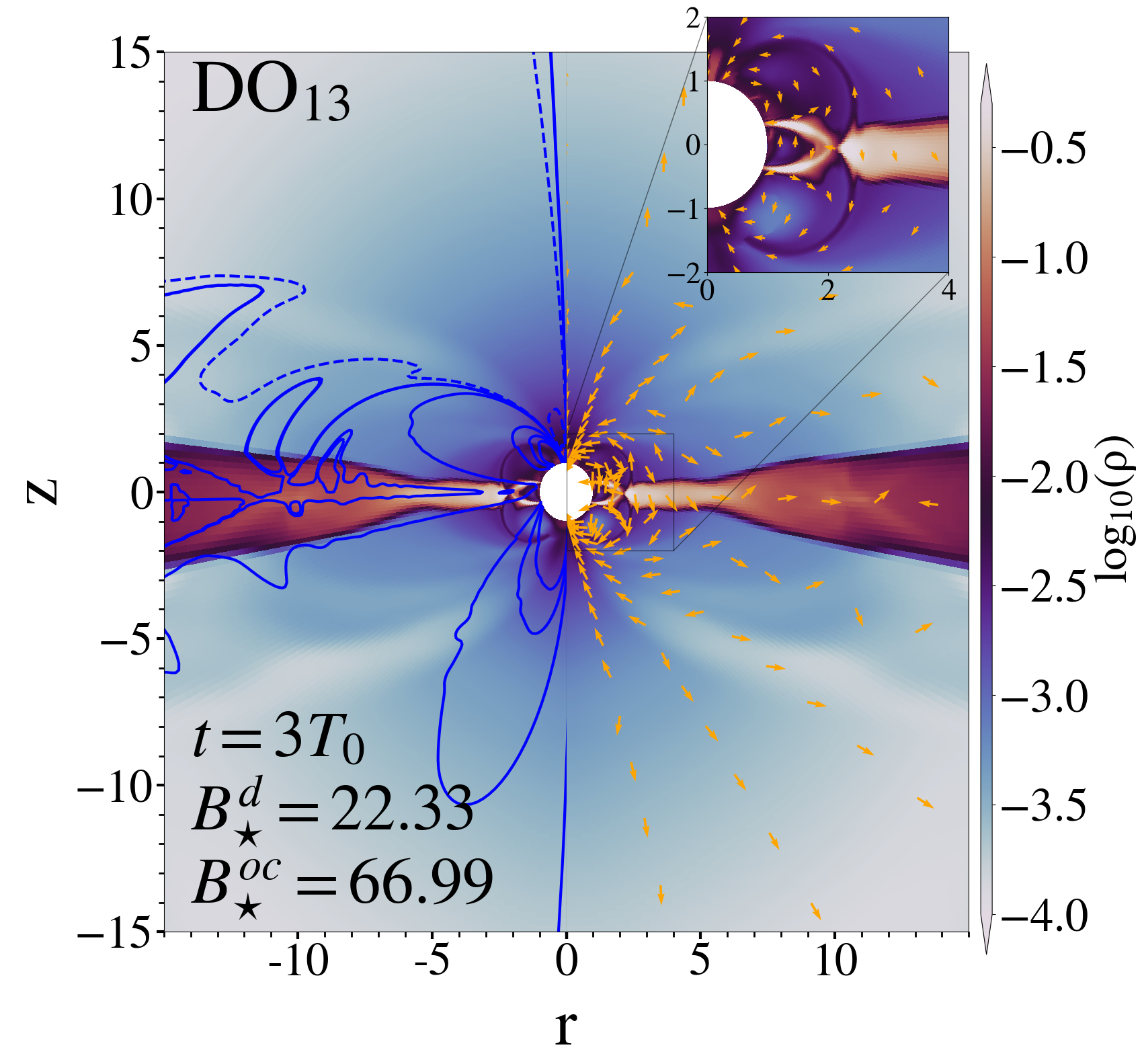}
   \end{subfigure}
   \hfil
   \begin{subfigure}{0.48\textwidth}
       \includegraphics[scale=0.20]{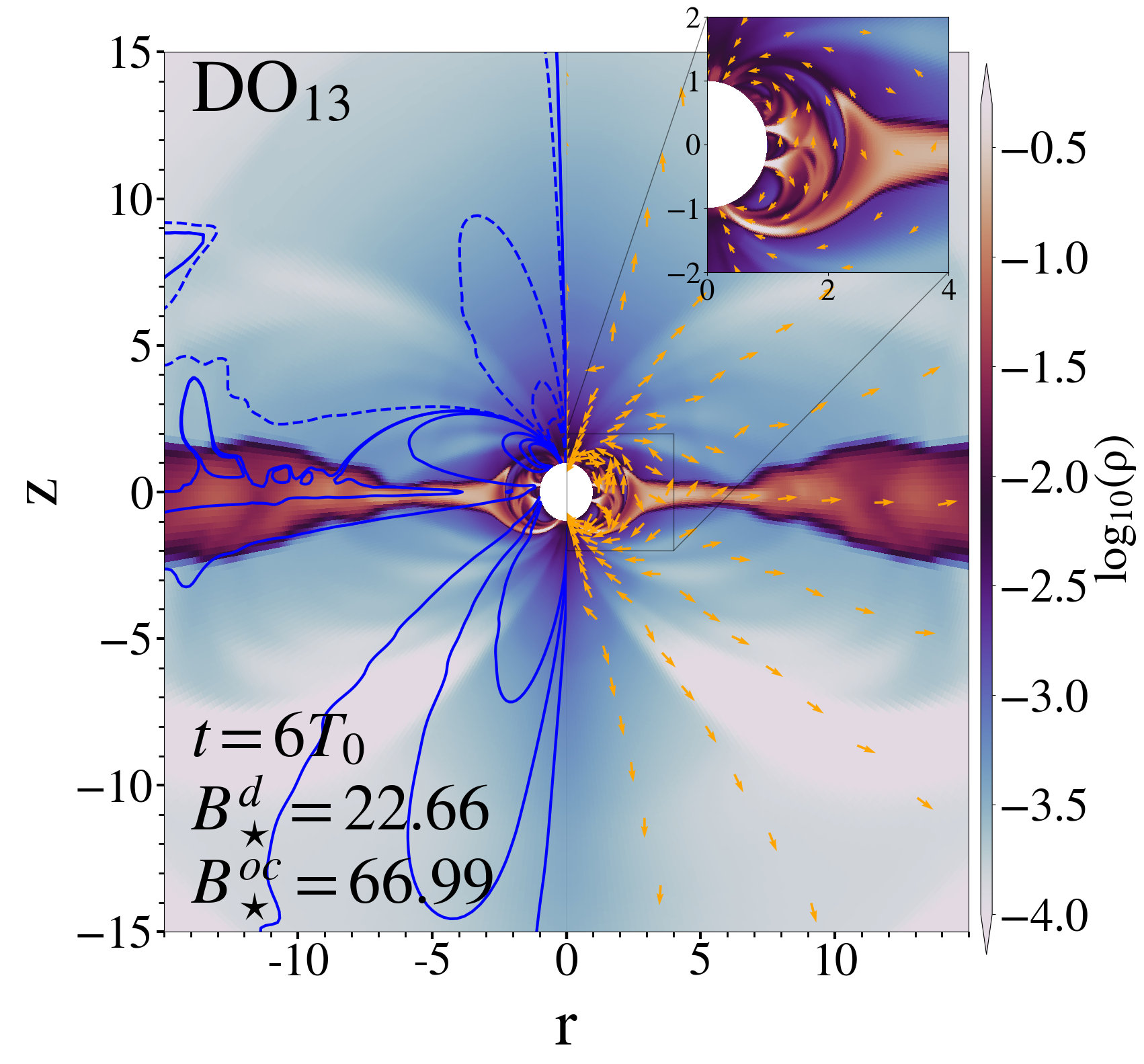}
   \end{subfigure}
    \caption{ Temporal evolution of the gas density (in logarithmic scale in units of $\rho_0$) for the model $\mathrm{DO}_\mathbf{13}$. Left panel corresponds to the time $t=3T_0$ and right panel corresponds to $t=6T_0$. The streamlines of the angular momentum fluxes $\mathbf{f}_B$ carried by the magnetic field are represented by the blue lines in the left side of each panel. While on the right side of each panel the velocity vectors are shown with yellow arrows. The inserts show a zoomed region of the density maps within $R<R_\mathrm{co}$.}
    \label{fig:Models_DO}
\end{figure*}

Figure \ref{fig:Models_DO} shows the temporal evolution of the gas density. At $t=3T_0$, a pair of well-defined funnels with a crab-claw topology forms, channeling material onto the star near the disk midplane. Additionally, two much fainter funnels are observed near the stellar poles. By $t = 6T_0$, multiple new funnels have developed, stacked within the original crab-claw structures, resulting in a total of six distinct funnels directing material onto the star at different latitudes. The latter result is a consequence of the topology of the magnetic field lines in the octupole (see third panel in Fig. \ref{fig:initial comparison}). We stress that the funnels at the top are thinner and more tenuous. The density contrast between the top and bottom of the funnels is also reflected in the coronal region, as dense asymmetrical arches form below the midplane. Finally, we note that the lower outer funnel resembles a dipole shape.

\begin{figure*}
    \centering
\begin{subfigure}{0.48\textwidth}
      \includegraphics[scale=.22]{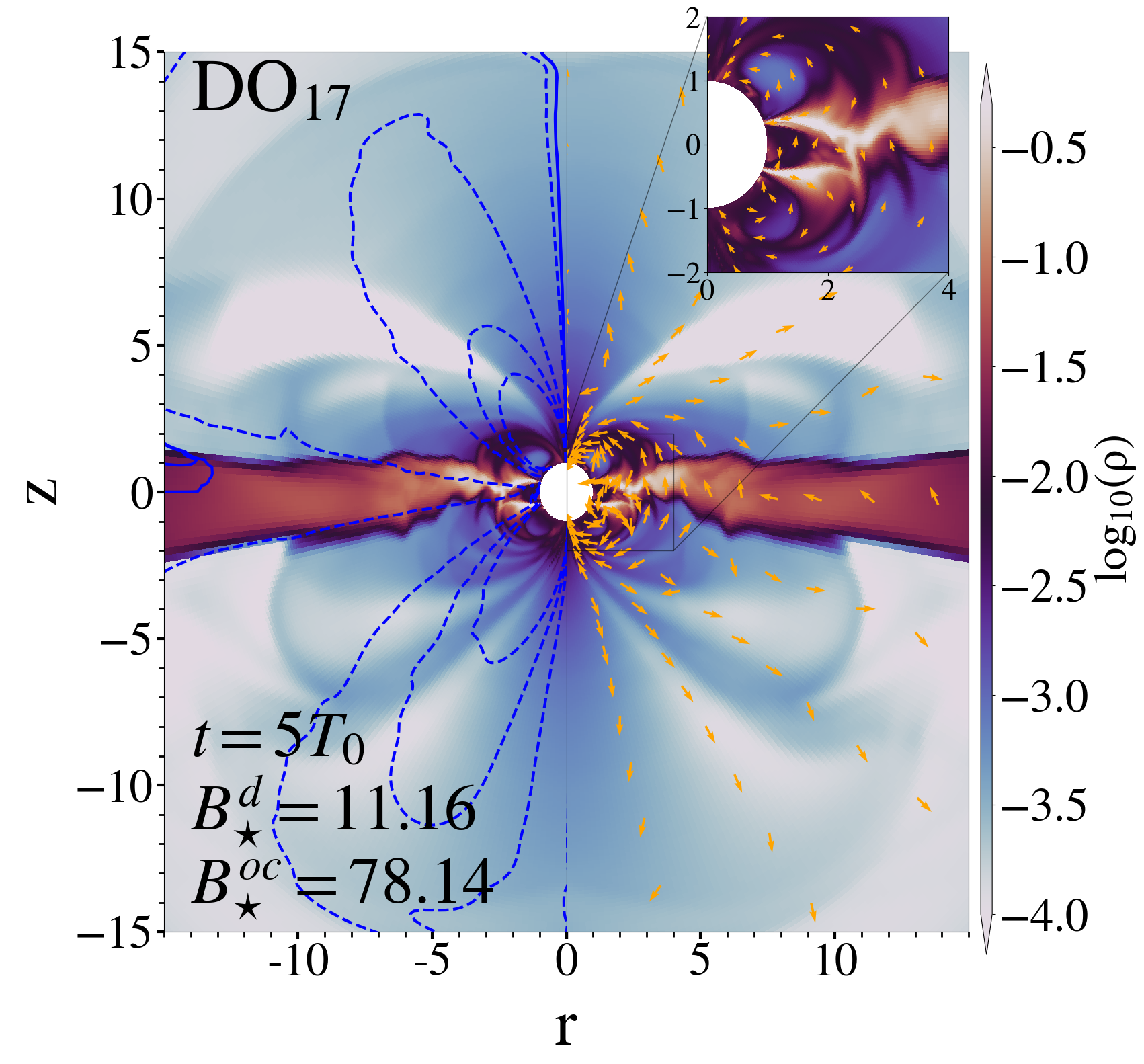}
   \end{subfigure}
   \hfil
   \begin{subfigure}{0.48\textwidth}
       \includegraphics[scale=0.22]{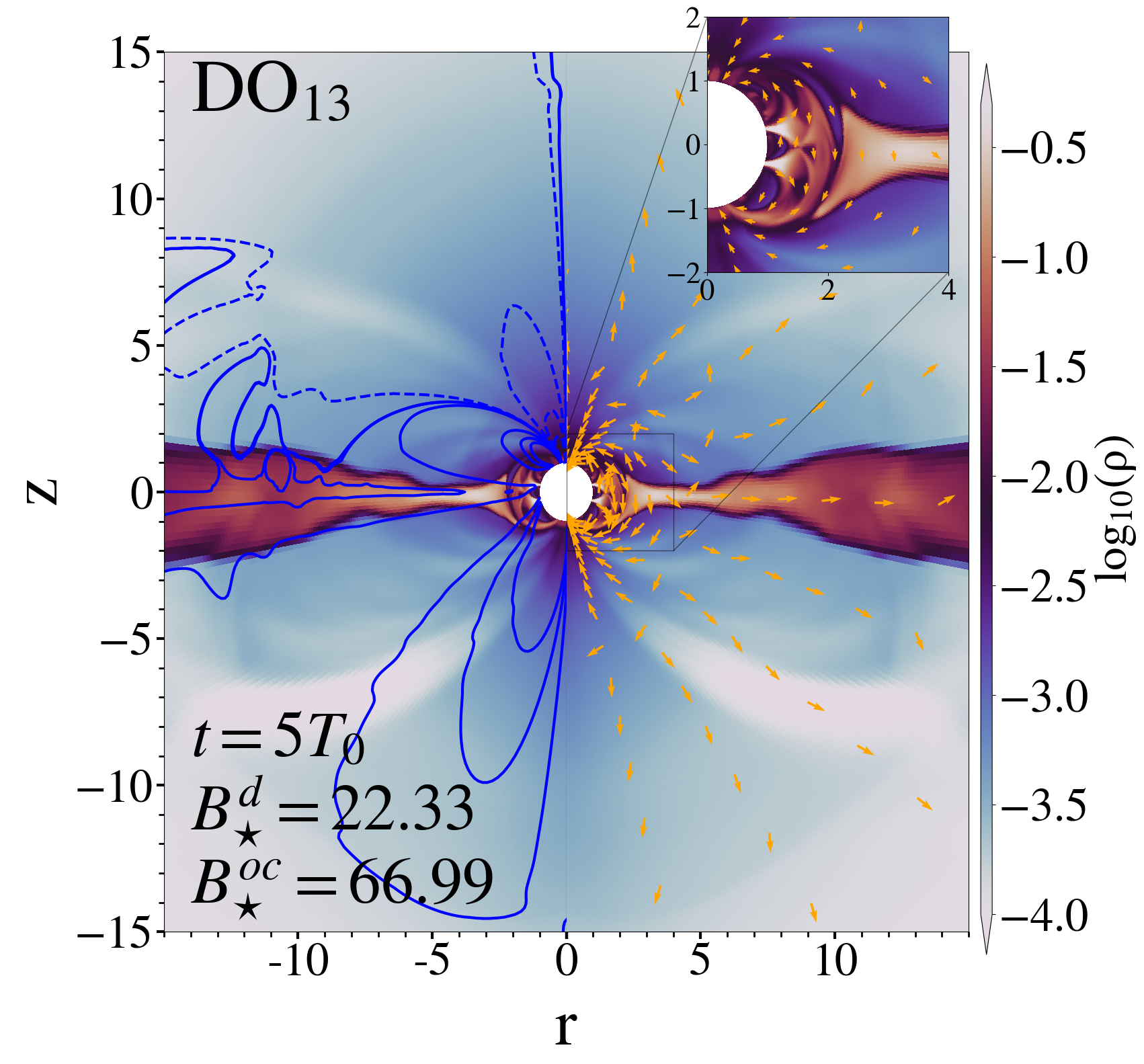}
   \end{subfigure}
   \caption{ Gas density comparison between dipolar plus octupolar models in logarithmic scale in units of $\rho_0$ at $t=5T_0$. Left panel correspond to the model $\mathrm{DO_{17}}$, at $t=5T_0$, and right panel is for the model $\mathrm{DO_{13}}$ at the same time of evolution. The blue lines and inserts are the same as in Fig. \ref{fig:Models_DO}. The gas flow is compared at $t=5T_0$, when a quasi-steady state has been reached in the mixed dipolar-octupolar configurations.}
\label{fig:comparison_DO}
\end{figure*}

On the other hand, for model $\mathrm{DO}_{\textbf{17}}$, we observe again the formation of multiple funnels following a pattern very similar to the magnetic field lines shown in the lower panel of Fig. \ref{fig:initial comparison}. In this case, the magnetic field strength of the octupolar component is seven times greater than that of the dipolar component. As a result, the accretion funnels above and below the midplane of the disk are both well defined. Nevertheless, 
the inclusion of the dipolar component introduces a twist in the region where each funnel originates, ultimately leading to an asymmetry between the upper and lower regions of the disk and corona (see left panel in Fig. \ref{fig:comparison_DO}).

In Fig. \ref{fig:comparison_DO} are  compared models $\mathrm{DO}_{\textbf{13}}$ and $\mathrm{DO}_{\textbf{17}}$ at $t=5T_0$. Far from the star ($R>10R_\star$), the disk structure is similar in both cases. However, for $R\leq10R_\star$ and in the regions where the funnels form, there is clearly a substantial difference. As the strength of the dipolar magnetic field increases (model $\mathrm{DO}_{\textbf{13}}$), the disk is compressed. Interestingly, the gas accumulates to a greater extent below the midplane, which may lead to a different accretion rate compared to the quasi-octupolar case ($\mathrm{DO}_{\textbf{17}}$). In both models, the perturbations in the corona are asymmetric with respect to the midplane (an effect that can also be inferred from the streamlines of angular momentum, see blue lines in Fig. \ref{fig:comparison_DO}), with  the model $\mathrm{DO}_{\textbf{17}}$ exhibiting a stronger perturbation.


\section{Discussion}
\label{sec:discussion}

\subsection{Observational motivation for the different magnetic configurations}

It has been suggested that the magnetic configuration on the surface of the stars may be a composition of dipolar, quadrupolar and even octupolar configurations \citep[e.g.,][]{Zanni2009, Cemeljic2019}. On the other hand, observations suggest that the magnetic field on the surface of stars may be more complex than the dipolar configuration usually considered in numerical studies. Preliminary spectropolarimetric results for the observations of one FS~CMa star, IRAS 17449+2320, show a strong magnetic field with a predominant dipolar configuration plus quadrupolar contribution. The estimated dipolar field strength is around $9000$ G, while the quadrupolar component is around $3000$ G (Bermejo-Lozano et al., in prep). For this reason, to model the corona of FS~CMa-type stars, it is reasonable to assume different configurations of the magnetic field on the stellar surface. 

\begin{figure}
    \centering
    \includegraphics[width=1.0\linewidth]{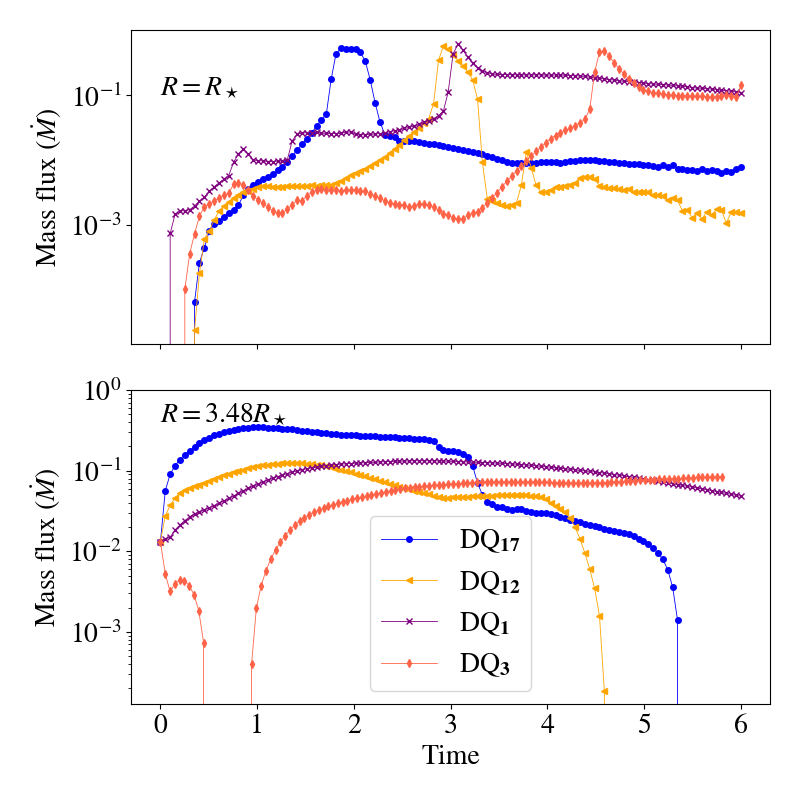}
    \caption{Time dependence of the mass flow rate for DQ models at $R= R_\star$ (upper panel) and at the truncation radius $R= 3.48R_\star$ (lower panel). 
    }    
    \label{fig:m_dot}
\end{figure}

\subsection{Twisted funnel flow accretion}

As we have shown in Section \ref{sec:results}, the accretion onto the star depends on the relative contributions of the multipolar components. 
Depending on the dominant magnetic polar strength in the configuration, accretion can occur through two funnel flows that are completely symmetric with respect to the disk midplane (purely dipolar configuration), or it can be drastically modified (see Subsection \ref{ssub:beyond} for a further discussion).

In a DQ configuration, regardless of the $B_\star^d/B_\star^q$ ratio, a single twisted funnel flow is formed, connecting the star below the disk midplane. 
That is, gas accretion onto the star occurs below the stellar equator. It is important to emphasize that the twisting of this single funnel flow becomes more pronounced as the quadrupole magnetic moment increases. Under this scenario, gas flows are generated above the disk midplane, which fall towards the stellar surface (see Fig. \ref{fig:dq_pollines}).  It is inferred that a twisted funnel flow can significantly change the mass flux of gas entering the star. Note that, in both the twisted funnel flow and the other cases studied here, the quasi-steady state is reached within a few star periods ($t\leq5T_\star$). With this in mind, in the next subsection we analyze the accretion of gas onto the star.

\subsubsection{Mass accreted in a disk}
\label{subsec:mass_accreted}
The mass flow rate crossing the whole polar range is:
 \begin{equation}
    \centering
    \dot{M}= -4\pi R^2\int_0^{\pi/2} \rho v_{R} \sin{\theta} \,d\theta.
\end{equation} 

Figure \ref{fig:m_dot}
shows the mass flow through the disk onto the surface of the star $R=R_\star$ and at the truncation radius $R_\mathrm{trunc}= 3.48R_\star$ given by Eq. (\ref{ec:rt_dip}), for the models DQ$_\mathbf{17}$, DQ$_\mathbf{12}$, DQ$_\mathbf{1}$ and DQ$_\mathbf{3}$.
In all four models, the magnitude of mass flux at $R=R_{\star}$ increases with the strength of dipolar component. In other words, when the intensity of the dipolar magnetic field increases, $\dot{M}$ also increases until it reaches a quasi-stationary value (which occurs in about 5 orbital periods of the star). From this figure, we can infer the dependence of $\dot{M}$ on the path the gas follows toward the star, which is constructed based on the initial magnetic configuration. For instance, for the models DQ$_\mathbf{1}$ and DQ$_\mathbf{3}$, $\dot{M}$ takes a very similar value ($\sim0.1$) at $t>4.5T_0$, which is, in fact, higher than it is for the other two models (where $\dot{M}\simeq10^{-3}$), with the quadrupole component making a larger contribution. The above argument can be confirmed from the funnel structures shown in Fig. \ref{fig:dq_pollines}. The funnels in models DQ$_\mathbf{1}$ and DQ$_\mathbf{3}$ exhibit a thinner structure, and the footpoint of these funnels is closer to the midplane of the disk. This latter feature is a consequence of the strong compression of the disk between $R\simeq R_\mathrm{trunc}$ and $R\simeq10R_\star$. Therefore, the distance traveled by each fluid element is shorter in these cases than in models where the funnel arch is increased, such as DQ$_\mathbf{17}$ and DQ$_\mathbf{12}$. 
  
Consider now the flux through the sphere with radius $R=3.48R_\star$ (that is, the truncation radius given by Ec. (\ref{ec:rt_dip}) for a pure dipolar configuration). A faster flow towards the star occurs
in the DQ$_\mathbf{17}$ configuration within the first three orbital periods. This is expected because, although a twisted funnel forms, the dominance of the quadrupolar magnetic component leads to a greater gas flow above and below the disk midplane
(see the first panel in Fig. \ref{fig:dq_pollines}). Nevertheless, for $t\geq3T_0$, the mass flow decreases in a stepwise manner, corresponding to episodic expansions in the length of the funnel flow (due to the expansion of the magnetic field lines). Something similar happens in the model DQ$_\mathbf{12}$ when $t\geq2T_0$.

\begin{figure}
    \centering
  \includegraphics[width=1.0\linewidth]{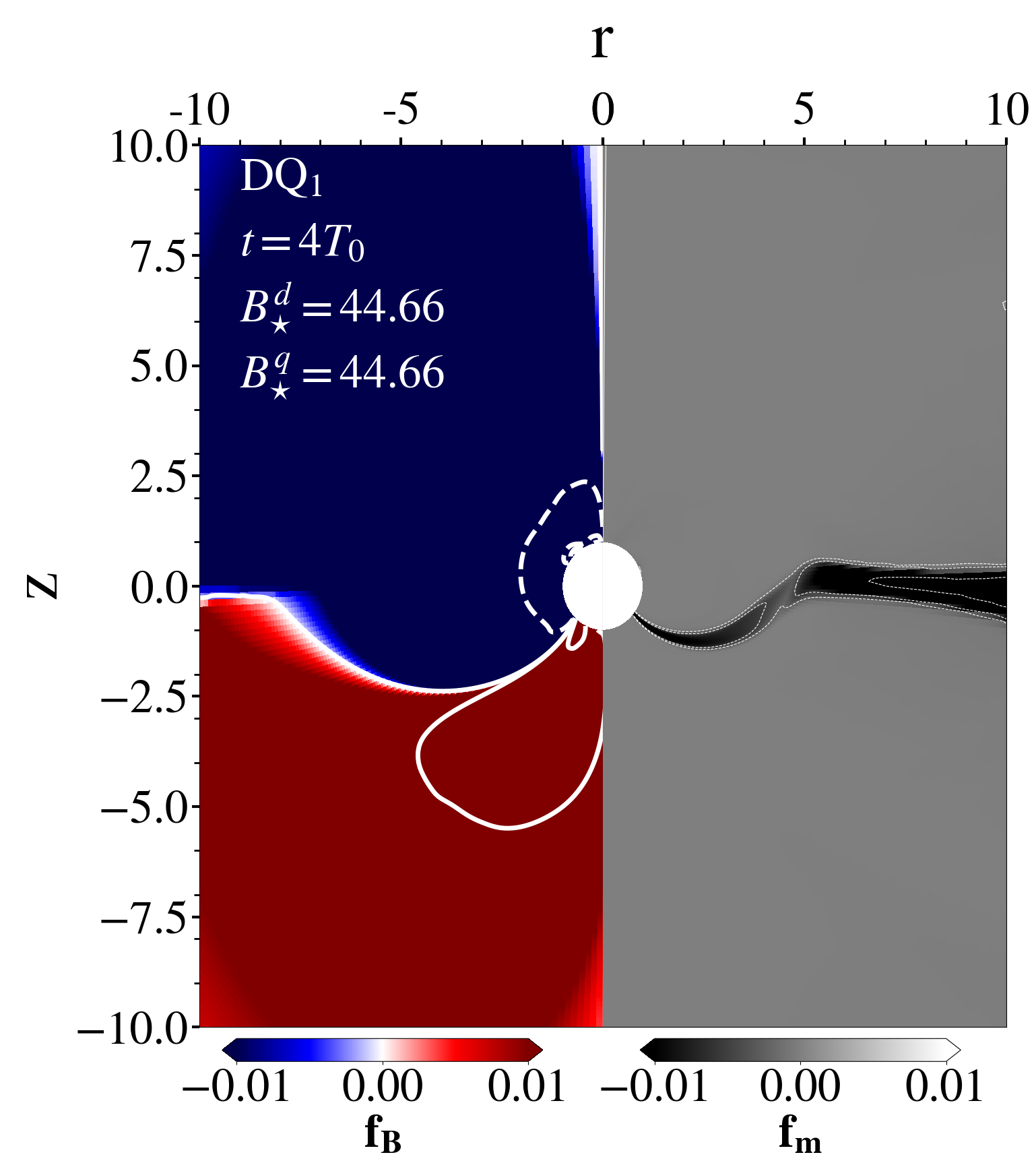}
    \caption{Maps of the angular momentum fluxes carried by the field $\mathbf{f}_B$ and the matter $\mathbf{f}_m$ (see Eqs. \ref{eq:fb} and \ref{eq:fm}). Note that the arc formed by the matter flux $\mathbf{f}_m$ (right-side) is contained within the region of negative angular momentum flux $\mathbf{f}_B$ through the magnetic field (left-side).}
    \label{fig:fluxes}
\end{figure}

On the other hand, when the dipolar magnetic moment is equal to or greater than the quadrupolar magnetic moment, as in the models DQ$_\mathbf{1}$ and DQ$_\mathbf{3}$, the mass flux toward the star remains quasi-stationary for $t>1T_0$, which means that the funnel is more stable and does not undergo strong radial expansion. The behavior in the model DQ$_\mathbf{1}$ (where the intensity of the dipolar and quadrupolar magnetic fields are equal) suggests that the angular momentum flux is mainly carried by the magnetic field. To verify this hypothesis, we consider 
the angular momentum fluxes carried by the field $\mathbf{f}_B$, and by the matter
$\mathbf{f}_m$ (see Eqs. \ref{eq:fb} and \ref{eq:fm}).
Figure \ref{fig:fluxes} shows the maps of $\mathbf{f}_B$ and $\mathbf{f}_m$ for model DQ$_\mathbf{1}$ at $t=4T_0$. It can be seen that the angular momentum flux driven by the field is more negative in the region where the funnel formation takes place (see left panel) compared to the angular momentum flux $\mathbf{f}_m$ driven by the matter (see right panel). This produces a radial shift of the funnel flow that inevitably results in a sharp drop in the mass flow $\dot{M}$ across the truncation radius (see lower panel in Figure \ref{fig:m_dot}).

\begin{figure}
 \centering
 \includegraphics[width=1.0\linewidth]{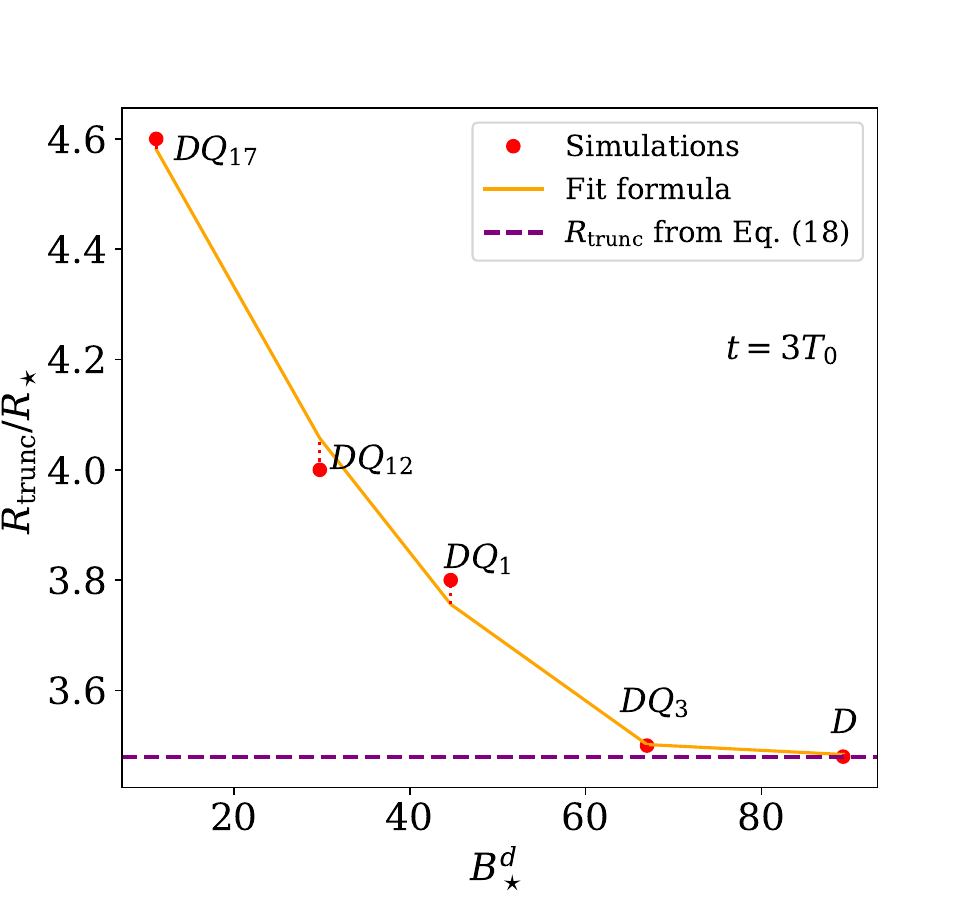}
     \caption{Truncation radius at $t=3T_0$ for DQ models. The intersection point of the dashed purple and solid orange lines represents the theoretical truncation radius value ($R= 3.48R_\star$) using the Eq. (\ref{ec:rt_dip}).}    
    \label{fig:R_T}
 \end{figure}

\subsubsection{Tracking the quadrupolar truncation radius}
\label{ssub:rt_fit}

As specified in Subsection \ref{subsec:rt}, the theoretical value of the truncation radius for the purely dipolar and octupolar cases is determined by Eqs. (\ref{ec:rt_dip}) and (\ref{ec:rt_oct}), respectively. However, due to the nature of the initial configuration of the magnetic field components, which implicitly depend on a poloidal configuration, it is not possible to analytically predict the value of the truncation radius for the quadrupolar case at the disk midplane.

To obtain an expression for $R_\mathrm{trunc}$ for the quasi-quadrupolar case, we calculate the truncation radius from the simulation results for each of the DQ models and insert it into function of the intensity of the dipolar component of the magnetic field. From these data we obtain the next fitting formula:
\begin{equation}
\frac{R_\mathrm{trunc}}{R_\star}=K_D\left(B_\star^d\right)^2-K_QB_\star^d+K_0
    \label{eq:ffit}
\end{equation}
where $K_D=2.37\times10^{-4}$, $K_Q=3.78\times10^{-2}$ and $K_0=4.97$, respectively.

Figure \ref{fig:R_T} shows the comparison of the fitting formula given in Eq. (\ref{eq:ffit}) and the simulation data. For reference, we have included the value of the purely dipolar case predicted by Eq. (\ref{ec:rt_dip}). We see that our formula adequately predicts the truncation radius when the quadrupolar component is considerably dominating.

\subsubsection{Beyond the dipolar funnel flow}
\label{ssub:beyond}

In the standard funnel flow formation, the gas falling onto the star follows the path defined by the magnetic field lines. In general, we found this behavior across all the magnetic configurations studied. However, the fact that the gas follows the magnetic field does not necessarily imply that a funnel flow is formed.
For instance, a purely quadrupolar configuration produces an accretion flow confined to the disk midplane, thereby preventing the lifthing of material above the disk midplane and formation of a funnel flow (see right panel in Fig. \ref{fig:ComparisonDQ}). Furthermore, we find multiple funnel formation for octupolar cases.

Note that even for a longer time evolution (which is beyond the scope of this work) in both the octupolar configurations and the other magnetic configurations analyzed in this study, the inflation of the magnetic field lines is supported by the differential rotation between the star and the gas, as well as the value of the magnetic diffusivity in the disk \citep[extended magnetosphere, see][and references therein for details]{Moranchel24}. Furthermore, in a dipolar magnetic configuration including background magnetic field, if the constraint of a thin disk is relaxed, accretion bursts, i.e. gas streaming directly from the truncation radius to the star induced by the interchange instability, can be found as well as one-sided outflows for stars with kilogauss magnetic fields. In such cases magnetospheric accretion develops \citep[see][for details]{GTR2024}. Therefore, the gas accretion rate and density structures (including the twisted funnels formed in DQ-models) can remain stable longer than reported here.

\section{Conclusions}
\label{sec:conclusions}

We have performed $2.5$-D MHD simulations of star-disk magnetospheric interaction, aimed at checking the gas density structure and the gas accretion process in massive stars with strong magnetic field characteristic in several Herbig Ae/Be stars and FS CMa stars. 

We have found that the gas accretion process is highly dependent on the symmetric (asymmetric) configurations of the magnetic field. We can classify accretion flows as follows:
\begin{itemize}
    \item [(i)] In a dipolar (symmetric) magnetic configuration, two funnel streams form, symmetric with respect to the disk's midplane.
    \item [(ii)] In a quadrupolar (symmetric) magnetic configuration, accretion is driven along the disk's midplane, which is flattened by the quadrupolar magnetic field lines. This process is accompanied by a widening of the disk due to the conical shape of the stream, with the base of the cone located near the truncation radius.
    \item [(iii)] In the case of a dipole plus quadrupole (asymmetric with respect to the disk midplane) magnetic configuration,  we find the formation of twisted funnel flows when the quadrupolar polar strength component dominates. These funnel streams connect with the surface of the star below the disk midplane.
    \item [(iv)] Finally, for the dipole plus octupole configuration, we find the formation of multiple funnels behind the truncation radius predicted by Eq. (\ref{ec:rt_oct}). Remarkably, the inclusion of the dipolar component produces an asymmetry in the funnel structure with respect to the disk midplane.
\end{itemize}
    
In addition, in all our models, we found density substructures in the corona that follow different patterns depending on the magnetic field configuration, with these structures being more pronounced in multipolar configurations.
Since a multipolar magnetic field configuration can be the result of a post-merger event, this suggests that, for a strong magnetic field such as in the case of several FS~CMa stars, the disk and corona may exhibit a complex density distribution as consequence of post-merger event.
In future work, we plan to investigate the cases where magnetic field configurations at the surface of the star are misaligned  with respect to the rotational axis $\Omega$. To do so, the full 3D models will be considered.

\begin{acknowledgements}
      We thank the referee for useful suggestions. The work of R.O.C. was supported by the Czech Science Foundation (grant 21-23067M). Computational resources were available thanks to the Ministry of Education, Youth and Sports of the Czech Republic through the e-INFRA CZ (ID:90254).
      M\v{C} acknowledges the Czech Science Foundation (GA\v{C}R) grant No.~21-06825X and the Polish NCN grant 2019/33/B/ST9/01564.
\end{acknowledgements}

%
   \bibliographystyle{aa} 
   \bibliography{references1} 
%

\appendix
\section{Vector potentials and flux functions}
\label{app:Apluspsi}
We assume that the star is a uniformly magnetized rotating sphere, on which flow the azimuthal currents $I(\theta)$.  The distribution of currents $I(\theta)$ will depend on the magnetic configuration so that if it is a purely dipolar magnetic field $I(\theta)=I_d \sin{\theta}$, in the case of a purely quadrupole magnetic field $I(\theta) = I_q \sin{\theta}\cos{\theta}$, while when it is a purely octupolar configuration we  have $I(\theta) = I_{oc} \sin{\theta}\,(5\cos^2{\theta}-1)$ \citep{Shakura91}. Since we use a simplified axisymetric model, the vector potential has only azimuthal component. Therefore, the potential vector for dipolar, quadrupolar and octupolar configurations reads
\begin{equation}
    A^{d}_\phi \equiv \frac{\mu_d}{R^2}\sin{\theta},
\end{equation}
\begin{equation}
    A^{q}_\phi \equiv \frac{3\mu_q}{4R^3}\sin{\theta}\,\cos\theta,
\end{equation}

\begin{equation}
    A^{oc}_\phi \equiv \frac{\mu_{oc}}{2R^4}\sin{\theta}\,(5\cos^2{\theta}-1).
\end{equation}

\noindent By definition, $\vec{B}=\nabla \times \vec{A}$ and, after some algebra, we can derive the components of the magnetic field.\\

The contours of the magnetic flux function
\begin{equation}
    \psi= A_\phi R \sin{\theta}
\end{equation}
trace the poloidal field lines. The magnetic flux functions for the dipolar, quadrupolar and octupolar fields are given by

\begin{equation}
    \psi^{d}= \frac{\mu_d \sin^2{\theta}}{R},
    \label{eq:psi_d}
\end{equation}
  
\begin{equation}
    \psi^{q}= \frac{3\mu_q}{4R^2}\sin^2{\theta}\cos{\theta},
    \label{eq:psi_q}
\end{equation}
and
 \begin{equation}
    \psi^{oc}= \frac{\mu_{oc}\sin^{2}\theta \,(5\cos^2\theta-1)}{2R^3}.
    \label{eq:psi_o}
\end{equation}

\end{document}